\documentclass[lettersize,journal]{IEEEtran}
\usepackage{graphicx,comment}
\usepackage[colorlinks=true, allcolors=blue]{hyperref}
\usepackage{soul}
\usepackage{amsmath}
\usepackage{amssymb}
\usepackage{algpseudocode}
\usepackage{enumitem}
\usepackage{mathrsfs}
\usepackage[table]{xcolor}
\usepackage{caption}
\usepackage{subcaption}

\usepackage{bbm}
\usepackage{algorithm}
\usepackage{tabularx}

\newtheorem{remark}{\textbf{Remark}}
\usepackage{fancyhdr}

\makeatletter
\let\ps@IEEEtitlepagestyle
\makeatother

\ifCLASSOPTIONcompsoc
  \usepackage[nocompress]{cite}
\else
  \usepackage{cite}
\fi

\ifCLASSINFOpdf
 
\else
  
\fi

\begin{document}

\title{Planning and Operation of Millimeter-wave Downlink Systems with Hybrid Beamforming}

\author{Yuan~Quan, Shahram~Shahsavari and~Catherine~Rosenberg,~\IEEEmembership{Fellow,~IEEE}
        
\thanks{Y. Quan and C. Rosenberg are with ECE department,  University of Waterloo (UW), Canada. S. Shahsavari is with Huawei Technologies R\&D, Canada. This work was done while he was a post-doctoral fellow at UW.}}


\maketitle

\vspace{-30pt}
\begin{abstract}
This paper investigates downlink radio resource management (RRM) in millimeter-wave systems with codebook-based hybrid beamforming in a single cell. 
We consider a practical but often overlooked multi-channel scenario where the base station is equipped with fewer radio frequency chains than there are user equipment (UEs) in the cell. In this case, analog beam selection is important because not all beams preferred by UEs can be selected simultaneously, and since the  beam selection cannot vary across subchannels in a time slot, this creates a coupling between subchannels within a time slot. None of the solutions proposed in the literature deal with this important constraint. The paper begins with an offline study that analyzes the impact of different RRM procedures and system parameters on performance. An offline joint RRM optimization problem is formulated and solved that includes beam set selection, UE set selection, power distribution, modulation and coding scheme selection, and digital beamforming as a part of hybrid beamforming. The evaluation results of the offline study provide valuable insights that shows the importance of not neglecting the constraint and guide the design of low-complexity and high-performance online downlink RRM schemes in the second part of the paper. The proposed online RRM algorithms perform close to the performance targets obtained from the offline study while offering acceptable runtime.

\end{abstract}
\vspace{-5pt}

\IEEEdisplaynontitleabstractindextext

\IEEEpeerreviewmaketitle

\section{Introduction}\label{sec:introduction}

Massive multi-input multi-output (MIMO) technology has emerged as a promising solution to address the increasing demand for high spectral and energy efficiencies in cellular networks. Moreover, millimeter wave (mmWave) bands, operating between 30 GHz and 300 GHz, have shown significant potential in delivering ultra-high data rates. Nevertheless, mmWave bands pose unique challenges, notably in terms of path loss and shadowing, which are more severe compared to microwave bands \cite{BF1}.


Beamforming (BF) plays a critical role in enabling wireless communications in mmWave systems. While fully digital BF (FDBF) has been widely recommended for microwave massive MIMO systems \cite{MIMO1}, its implementation in mmWave bands is costly and power-intensive due to the requirement of a separate radio frequency (RF) chain for each antenna \cite{hybrid-survey}. To mitigate the deployment cost and power consumption of FDBF, several hybrid BF (HBF) architectures are proposed, where the BF processing is divided between baseband digital and RF analog domains \cite{hybrid-survey}. These architectures utilize a large number of antennas with significantly fewer RF chains at the base station (BS) and user equipment (UE). HBF supports multiple (but limited) concurrent data streams to serve multiple UEs in the same time-frequency resource block and is more cost-effective and power-efficient than FDBF. Recent works have highlighted the critical role of HBF for future large-scale antenna systems, showing its importance from mmWave bands to terahertz frequencies and intelligent reflecting surface technology (e.g., \cite{HybridAdded1, HybridAdded2}).

\begin{figure}
    \centering
    \includegraphics[width=70mm]{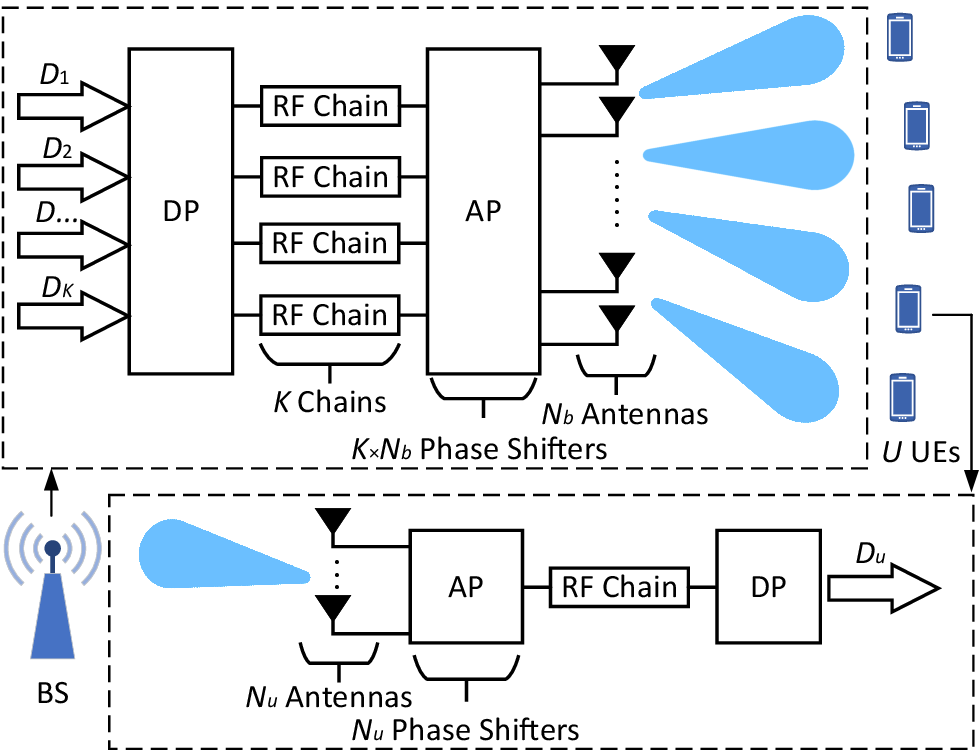}
    \caption{Fully connected HBF architecture at the BS and UE 
    }
    \label{fHBF}
    \vspace{-10pt}
\end{figure}

In this paper, we consider the downlink of a single small cell operating in mmWave band using HBF. The BS is equipped with $K$ RF chains connected to an antenna array while the UEs have a single RF chain connected to an antenna array. We consider a fully connected HBF architecture, depicted in Fig.~\ref{fHBF}, where BF is divided into baseband digital processing followed by RF analog processing. A limited number of RF chains connect the digital processor (DP) and the analog processor (AP). 
The baseband DP can be used for digital BF (DBF) which is also known as precoding. 
The AP is used to implement analog BF (ABF) in the RF domain by employing a group of phase shifters as illustrated in Fig.~\ref{fHBF}. We will consider both the cases with and without DBF in the following.



While DBF and ABF can be optimized as illustrated in \cite{MMWAVE1}, it might be more practical to choose ABF vectors (i.e. the vector of the phase shifts applied at the BS and UEs) from pre-designed ABF codebooks \cite{ConferenceVersion} and to pre-select a precoding technique a priori. In this paper, we use a codebook-based HBF scheme which is also promoted by 5G NR \cite{CE1} and zero-forcing for DBF \cite{ECSIXiaomeng}. The ABF codebooks can be designed in a planning phase with various techniques \cite{CODEBOOK1,Codebook-tork}.

The operation of a cellular system involves the following procedures performed at different time scales \cite{ECSIAhmed,ECSIGuosen,ECSIShiwen,ECSIXiaomeng}: first, \textit{beam alignment} (BA) determines the ABF vectors \footnote{We use the term ABF vector and beam interchangeably.}  for each UE \cite{BA2,BA5}. We match each UE with one pair of ABF vectors (one from the BS codebook and one from the UE codebook). We will also consider the case where each UE is matched with multiple pairs of preferred beams as an extension in Section~\ref{sec:offline}. Second, \textit{channel estimation} (CE) estimates the effective channel state information (ECSI) of each UE periodically in time and frequency for the preferred beam pair(s) of each UE. Third, \textit{radio resource management} (RRM) is performed periodically using the most recent ECSI for each UE. Fourth, \textit{data transmission} takes place. Note that BA and CE provide inputs to RRM but they are not in the scope of the paper focus.

Our research mainly focuses on the RRM in a multiple-channel scenario with fewer RF chains than UEs in the cell (i.e., $K<U$ where $U$ is the number of UEs). This scenario is overlooked by the literature although it is highly practical. In this case, analog beam selection is important because not all beams preferred by UEs can be selected simultaneously, and since the  beam selection cannot vary across subchannels in a time slot, this creates a coupling between subchannels within a time slot. Hence, the methodologies developed in the literature for the case $K\geq U$ cannot be directly extended to our scenario since they all focus on a single subchannel. In general, RRM comprises:

\begin{enumerate}[leftmargin=*]
\item \textit{Power allocation}: The approach taken to share the BS transmit power among the frequency subchannels in a given time slot. In this paper, we consider equal power allocation among frequency subchannels and leave power allocation optimization for future research.
\item \textit{Beam selection}: Unlike most of the papers in prior art, we consider multiple frequency subchannels instead of assuming only one subchannel.  Since ABF is realized by phase shifters, the same set of ABF vectors must be considered for all the subchannels in a time slot.  Moreover, the number of BS ABF vectors that  can be selected in every time slot has to be less than the number of RF chains which might be less than the number of users. Therefore, a subset of BS beams might have to be selected for each time slot. Unlike most of the papers in prior art, we consider multiple frequency subchannels instead of assuming only one subchannel. This constraint is essential since ABF is realized by phase shifters, hence, the same set of ABF vectors must be considered for all the subchannels at a time slot of the system, i.e., applied ABF vectors cannot vary over the subchannels. This constraint can complicate RRM by preventing it from taking a per-subchannel approach as all the subchannels should be considered jointly in a time slot. To circumvent this challenge, several prior studies (e.g., \cite{ECSIAhmed,ECSIGuosen}) assumed that the number of RF chains at the BS is not smaller than the number of UEs. This assumption implies that all the preferred beams can be selected at every time slot making beam selection trivial and RRM can then be performed on a per-subchannel basis. 
\item \textit{User selection}: Given the selected BS beam set for a time slot, for each subchannel in that time slot, we need to select a UE set containing at most one UE for each beam in the beam set. Note that only the UEs whose preferred beam is in the selected beam set can be chosen.

\item \textit{Digital BF}: To enhance the performance, one can apply MIMO precoding to the data streams of the selected UEs in a given PRB using DBF. In this paper, we consider zero-forcing DBF (ZF-DBF) as well as scenarios without DBF where only ABF is present.  

\item \textit{Power distribution (PD)}:  The way that the power allocated to a subchannel is distributed among the selected UEs for that subchannel. There are various approaches for PD, from simply sharing the power equally among all selected UEs to distributing power to maximize a system utility function. We consider both scenarios in this paper.
\item \textit{Modulation and coding scheme (MCS) selection}: The outcome of the previous steps is an estimate of SINR per selected UE which is used to select an appropriate MCS. 
\end{enumerate}
Our research goal is to propose online RRM schemes that strike a balance between high performance and low complexity. To achieve this objective, we conduct an offline planning study to investigate the impact of various RRM procedures mentioned above as well as different system parameters such as number of RF chains, number of UEs, and BF codebook size. We formulate an offline joint RRM optimization problem that includes beam and user selection, DBF, PD, and MCS selection. Our offline study produces valuable insights into system parameter choices, as well as the impact of PD and DBF, which we apply in the second part of the paper to develop low-complexity and high-performance online DL RRM schemes. Our proposed online algorithms are heuristics that handle  optimized PD with optional ZF-DBF. The main contributions of this paper are summarized below.
\begin{itemize}[leftmargin=*]
  \item We assess the downlink performance of a small cell massive MIMO system with codebook-based HBF operating in mmWave band, where each UE is mapped to a single preferred beam pair. In contrast to existing literature, our study considers a setting in which the number of RF chains at the BS may be smaller than the number of UEs, and multiple frequency subchannels are in use. Consequently, not all preferred BS beams can be selected at the same time, and the selected beam set cannot be changed from one subchannel to another within a time slot. To address this constraint, we formulate a joint optimization problem, which includes beam selection, user selection, optional ZF-DBF, PD and MCS selection, on a per-time slot basis rather than a per-subchannel basis. Our analysis shows that neglecting the beam-set constraint and solving the problem on a per-subchannel basis can overestimate the performance by 20\%.  
  
  \item We conduct a comprehensive planning study in which we propose offline solution methods to the formulated optimization problems for equal PD (EPD) and optimized PD (OPD) cases. Furthermore, we consider two scenarios for each case: without DBF (N-DBF) and with ZF-DBF. For the EPD case, we obtain the optimal solution in both scenarios. For the OPD case, we solve the problem quasi-optimally in the ZF-DBF scenario and obtain locally optimal solutions in the N-DBF scenario. 
  
  \item We extend our formulations and solutions to the case where multiple beam pairs are selected for each UE during BA.

  \item We provide engineering insights based on the numerical evaluations obtained from our planning study. For instance, we show that ZF-DBF can improve the performance by 32\% compared to N-DBF. Moreover, we find that OPD can provide 22\% performance increase compared to EPD. Additionally, our study reveals that selecting two beam pairs per UE does not provide a large performance gain if the number of RF chains is small. Besides, we study the impact of the number of RF chains and the size of the ABF codebooks on the performance. 

  \item We design low-complexity online heuristic RRM schemes for the OPD case with and without ZF-DBF for the case of one beam pair per UE (the case of two pairs is left for further study). Our heuristics can provide good performances compared with the performance benchmarks obtained in the planning study. The numerical results show that the performance of our best online heuristic is at least 92.3\% of the best offline benchmark performance. 


\end{itemize}

The rest of the paper is organized as follows. Section~\ref{sec:literature}  reviews  the related works. Section~\ref{sec:sysmodel} describes the system model.  Section~\ref{sec:offline} provides our planning (offline) study. In Section~\ref{sec:online}, we propose online heuristic RRM schemes for different scenarios. Section~\ref{sec:conclusion} concludes the paper.
\vspace{-5pt}
\section{Related Work} \label{sec:literature}
In this section, we review the prior studies in the literature related to DL RRM with codebook based HBF. 
Unlike our sequential approach where preferred ABF vectors are chosen in BA prior to RRM, several prior studies (e.g., \cite{CHANNELS1,SINR1,ECSIQianrui}) jointly optimize RRM with ABF. However, besides high computational burden, such optimizations require per-UE full CSI (i.e., MIMO channel matrix) which is typically difficult to obtain in practice especially for systems with frequency division duplexing (FDD) \cite{marzetta2016fundamentals}. In our approach, we select the ABF vectors for each UE during BA before RRM. This in turn reduces the complexity of BF and simplifies the channel estimation by reducing the dimension of the effective channels (from matrix to scalar). 

The problems studied in \cite{ECSIAhmed,ECSIGuosen,ECSIShiwen,ECSIXiaomeng} are similar in nature to the one we investigate in this paper. Additionally, BA is used to choose the ABF vectors for each UE in all of these studies. To clarify the distinctions, Table~\ref{RefTab} provides a list of the main system characteristics and assumptions considered in \cite{ECSIAhmed,ECSIGuosen,ECSIShiwen,ECSIXiaomeng} and our study. In this table, `\textit{ConB}' refers to the practical beam selection constraint discussed in Section \ref{sec:introduction}, i.e., the ABF vectors cannot vary in frequency domain within a time slot. Next, we elaborate on the novelties and distinctions.

In \cite{ECSIAhmed} and \cite{ECSIGuosen}, it is assumed that $K \geq U$ where $K$ is the number of RF chains at the BS and $U$ is the number of UEs. Consequently, beam selection is not necessary as all the preferred beams (ABF vectors) can be selected at each time slot. Nevertheless, it is not typical for a small cell BS to be equipped with a large number of RF chains in practice due to the high cost and complexity, hence, $K < |\mathcal{B}_p| \leq U$ can occur especially in crowded networks, where $\mathcal{B}_p$ is the set of the indices of all the preferred BS beams. Unlike \cite{ECSIAhmed} and \cite{ECSIGuosen}, we study RRM in a more general case where $K$ can be smaller than $|\mathcal{B}_p|$ and $U$. In this case, not all the preferred beams can be selected and beam selection must be considered. 

References \cite{ECSIShiwen} and \cite{ECSIXiaomeng} allow for $K < |\mathcal{B}_p|$. However, they restrict their studies to one frequency subchannel which is equivalent to RRM on a per-subchannel basis, implying the ignorance of the beam selection constraint.  We will show that the performance is significantly over-inflated when ignoring ConB and allocating the resources on a per-subchannel basis. That is the reason unlike \cite{ECSIShiwen} and \cite{ECSIXiaomeng}, we consider multiple frequency subchannels alongside beam selection constraint ConB, making the evaluations more practical. 

One of our objectives is to investigate the impact of PD and DBF on performance. Among the similar papers listed in Table \ref{RefTab}, only \cite{ECSIXiaomeng} adopts OPD and the rest consider EPD. To perform PD, \cite{ECSIXiaomeng} suggests using water-filling (WF) algorithm based on data rates computed by Shannon capacity formula. Note that while Shannon formula is commonly used, it may lead to misleading insights and wrong conclusions. In our research, we adopt a piecewise constant rate function modelling the existing practical MCSs rather than using Shannon capacity formula. In that case, WF algorithm is not the optimal solution \cite{MCS1}, \cite{MCS2} for PD. 

References \cite{MMSEDBF1} and \cite{MMSEDBF2} generate the DBF vectors with an iterative weighted sum-minimum mean square error algorithm, which introduces some complexity to both offline and online RRM schemes. 
Instead, we adopt ZF algorithm as the DBF technique similar to  \cite{ECSIXiaomeng}, \cite{ECSIAhmed} and \cite{ECSIGuosen}. 


Fairness is an important aspect of RRM. Among the similar papers listed in Table \ref{RefTab}, except \cite{ECSIXiaomeng}, the rest employ sum rate as the objective in their optimizations and evaluations. Even in \cite{ECSIXiaomeng}, only the proposed online algorithm is based on proportional fairness (PF). The derived performance upper-bound and numerical evaluations are based on sum rate. In this paper, We consider PF in the RRM optimization problem as well as when designing online heuristic algorithms. 

\begin{table}
    
\centering

\caption{Main system characteristics in similar studies from literature (CHN: channel, TS: time slot, FS: fairness, RF: rate function, OPT: optimized, ConB: beam selection constraint)}
\footnotesize
\setlength{\tabcolsep}{1.8mm}{
\begin{tabular}{|c|c|c|c|c|c|c|c|c|} 

\hline

\rowcolor{gray} & CHN & TS & $K < U$ & ConB & PD & DBF & FS & RF\\
\hline

\cellcolor[HTML]{D3D3D3} \cite{ECSIAhmed} & 1 & 1 & $\times$ & $\times$ & EPD & ZF & $\times$ & Sh \\
\hline

\cellcolor[HTML]{D3D3D3} \cite{ECSIGuosen} & 1 & 1 & $\times$ & $\times$ & EPD & ZF & $\times$ & Sh \\
\hline

\cellcolor[HTML]{D3D3D3} \cite{ECSIShiwen} & 1 & 1 & \checkmark & $\times$ & EPD & $\times$ & $\times$ & Sh \\
\hline

\cellcolor[HTML]{D3D3D3} \cite{ECSIXiaomeng} & 1 & $\geq 1$ & \checkmark & $\times$ & WF & ZF & PF & Sh \\
\hline

\cellcolor[HTML]{D3D3D3} Ours & $\geq 1$ & $\geq 1$ & \checkmark & \checkmark & OPT & ZF & PF & MCS \\
\hline

\end{tabular}}

\label{RefTab}
\vspace{-10pt}
\end{table}

\section{System Model} \label{sec:sysmodel}

We consider the DL of a single small cell
massive MIMO system operating in a mmWave band. The system is based on the fully connected HBF architecture (shown in Fig.~\ref{fHBF}) where the BS is equipped with $N_b$ antennas, $K$ RF chains and $N_b \times K$ phase shifters. Each RF chain is connected to all $N_b$ antennas by $N_b$ phase shifters. We use $\mathcal{U}$ and $U$ to denote the set and the number of UEs, respectively. Typically, we have $K\ll N_b$. In addition, $K$ can be smaller than $U$ in practice. We assume that each UE has one RF chain connected with each of its $N_u$ antennas by $N_u$ phase shifters.

\vspace{-10pt}
\subsection{Frame Structure and Overview of Operational Processes}

\begin{figure}
    \centering
    \includegraphics[width=60mm]{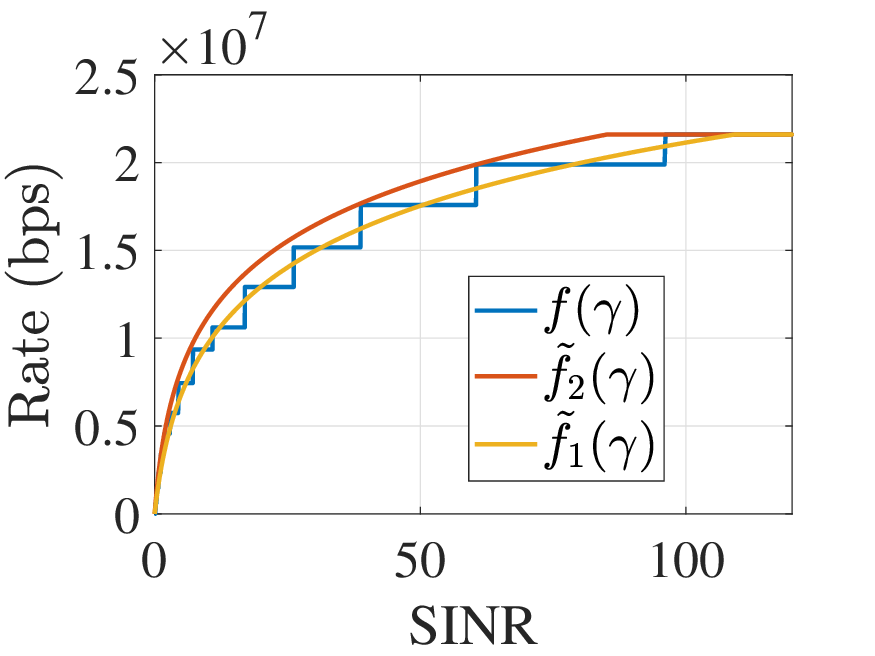}
    \caption{An example of piece-wise constant rate function $f(\gamma)$ and two convex approximations $\tilde{f}(\gamma)$, a tight one and an upper bound (see Section~\ref{subsec:ratefunc})}
    \label{fMCS}
\vspace{-10pt}
\end{figure}

The time-frequency resource unit for RRM is called a physical resource block (PRB). The time span of a PRB is denoted by $\delta_T$, which is the duration of a system time slot. The bandwidth of a PRB is  $\delta_F$, which is the bandwidth of a frequency subchannel. 


We assume that the channel state information (CSI) is obtained by CE periodically, i.e., at every \emph{reporting block} (RBL) corresponding to $N_F \times N_T$ ($N_F$ in frequency and $N_T$ in time) PRBs. We assume that one set of channel coefficients is reported for each UE within an RBL. Note that the size of RBL is an input to the RRM and can be as small as one PRB (i.e. $N_T=N_F=1$). The system bandwidth, denoted by $BW$, is divided into $Q$ RBLs in frequency. We define a \textit{mega block} (MB) as the group of $Q$ RBLs in the frequency domain (this is necessary to ensure the beam set constraint per time slot, please see later). In each MB, the DL operation is performed in three phases. The first phase is CE which is performed simultaneously at the beginning of each RBL. In the second phase, the estimated CSI is used to perform RRM for all the remaining PRBs in all the $Q$ RBLs of the MB. In the third phase, transmissions to the selected UEs are performed. 

\vspace{-10pt}
\subsection{Beam Alignment} \label{BA}
We assume that there are pre-defined ABF codebooks at the BS and the UE for BA. 
The BS codebook $\mathcal{C}_b$ and UE codebook $\mathcal{C}_u$ are defined as $\mathcal{C}_b=\{\mathbf{w}_j \in \mathbb{C}^{N_b \times 1}:\lVert \mathbf{w}_j \lVert ^2 =1,j=1,...,B_b\}$ and $\mathcal{C}_u=\{\mathbf{v}_j(u) \in \mathbb{C}^{N_u \times 1}:\lVert \mathbf{v}_j(u) \lVert ^2 =1,j=1,...,B_u\}$ respectively. In the following, we assume that all UEs use the same codebook and remove the index $u$.
 $B_b$ and $B_u$ are the ABF codebook sizes of BS and UE, respectively. 
 The objective of BA is to determine the best ABF vector pair(s) to connect the BS to each UE. This process is done periodically and, in the following, we consider RRM within one such period, i.e., the  preferred beams for each user do not change.
 We will first consider the case where a single pair is selected per UE and then, in Section~\ref{MultipeBeamPair}, we will consider the case of two pairs of ABF vectors per UE. The  RRM schemes that we will describe later in the paper are agnostic to the BA method. Indeed various methods can be used for BA (e.g., \cite{BA2}, \cite{BA5}, \cite{BA1}, \cite{BA3}). Specifically, we assume that beam pair $(\mathbf{w}_u^*,\mathbf{v}_u^*)$ is selected for UE $u$ during the BA process.

We define $\mathcal{C}_b^\star= \{\mathbf{w}_u^*\}_{u \in \mathcal{U}}$ as the set of all preferred ABF vectors at the BS. Additionally, we define $\mathcal{B}_p=\{b_1,...,b_{|\mathcal{C}_b^\star|}\}$ as the set of indices corresponding to the ABF vectors in $\mathcal{C}_b^\star$. 
After BA, the BS knows which ABF vector to use for transmission to UE $u$ and UE $u$ knows which ABF vector to use for reception if selected.
\vspace{-10pt}
\subsection{Effective Channel Estimation} \label{subsec:ECSI}
At the beginning of an MB,  the effective signal and interference channels are estimated for each UE simultaneously for each RBL of that MB. Specifically, during CE for RBL $q$, we obtain $g_{q,u,u}^\text{eff}$ for all $u$ and $g_{q,n,u}^\text{eff}$ for all $u$ and $n \neq u$, where $g_{q,u,u}^\text{eff} \in \mathbb{C}$ is the effective signal channel between UE $u$ and the BS in  RBL $q$ (it is the same for each PRB in the RBL) and $g_{q,n,u}^\text{eff} \in \mathbb{C}$ is the effective interference channel between the BS and UE $u$ due to the BS transmission to UE $n \neq u$ in any PRB of RBL $q$. We have $g_{q,u,u}^\text{eff}=\left({\mathbf{v}^*_u}\right)^H \mathbf{G}_{q,u}\left(\mathbf{w}_u^*\right)$ and $g_{q,n,u}^\text{eff}=\left({\mathbf{v}^*_u}\right)^H \mathbf{G}_{q,u}\left(\mathbf{w}_n^*\right)$ where $\mathbf{G}_{q,u}\in \mathbb{C}^{N_u \times N_b}$ denote the channel matrix between the BS and UE $u$ in RBL $q$. 
    We refer the reader to \cite{CE1}  for more details on CE. We define ECSI of UE $u$ at RBL $q$ as $\mathbf{g}^\text{eff}_{q,u} = [g_{q,1,u}^\text{eff},...,g_{q,U,u}^\text{eff}]^T \in \mathbb{C}^{U \times 1}$. Note that  ECSI captures the impacts of the propagation channel and ABF collectively. The RRM techniques that we will introduce later are independent of the CE method. For the purposes of this paper, we posit the perfection of CE and defer the exploration of CE inaccuracies to future research.


\vspace{-10pt}
\subsection{Digital Beamforming} \label{subsec:DBF}
DBF can improve the performance of HBF by providing more control on the interference. In this paper, we consider two cases: one with no DBF in which only ABF is present and another case with DBF. We adopt the well-known zero-forcing scheme whenever DBF is present. This type of precoding removes the interference among the UE data streams after ABF. In Section~\ref{sec:offline}, we will exploit the absence of interference to convexify and solve the  RRM optimization problem for the case with DBF.

Specifically, suppose that a set of beams, denoted by $\ell$ is selected in a time slot.  Furthermore, assume that an appropriate UE set $z_\ell=\{u_1,...,u_{|z_\ell|}\}$ is chosen for RBL $q$ in that time slot. At this point, we assume that a selection has been done for the beam set and the UE set. We will elaborate on how to perform beam set and UE set selections in Section~\ref{sec:offline} for the offline study and Section~\ref{sec:online} for the online study. Indeed, note that the main outputs of the RRM problems and algorithms that are the focus of this paper are the beam and user selections.

We define the ZF precoding matrix as $\mathbf{F}_q^{\text{ZF}} (z_\ell)=[\mathbf{d}^{\text{ZF}}_{q,u_1}(z_\ell),...,$ $\mathbf{d}^{\text{ZF}}_{q,u_{|z_\ell|}}(z_\ell)]$, where $\mathbf{d}^{\text{ZF}}_{q,k}(z_\ell)=\frac{\mathbf{a}_{q,k}(z_\ell)}{\lVert \mathbf{F}^{\text{ABF}}(z_\ell)\mathbf{a}_{q,k}(z_\ell) \lVert_F}$ for all $k \in z_\ell$, and $\mathbf{F}^{\text{ABF}}(z_\ell)=\big[\mathbf{w}_{u_1}^*,...,\mathbf{w}_{u_{|z_\ell|}}^*\big]$ is the ABF matrix including the selected ABF vectors for UE set $z_\ell$. We define $[\mathbf{a}_{q,u_1}(z_\ell),...,\mathbf{a}_{q,u_{|z_\ell|}}(z_\ell)]=\mathbf{G}^\text{eff}_q(z_\ell)^H [{\mathbf{G}^\text{eff}_q(z_\ell)} \mathbf{G}^\text{eff}_q(z_\ell)^H]^{-1}$ where $\mathbf{G}^\text{eff}_q(z)=[\mathbf{g}^\text{eff}_{q,u_1}(z_\ell),...,\mathbf{g}^\text{eff}_{q,u_{|z_\ell|}}(z_\ell)]^T \in \mathbb{C}^{|z_\ell|\times |z_\ell|}$, where $\mathbf{g}^\text{eff}_{q,u}(z_\ell)$ is the subvector of $\mathbf{g}^\text{eff}_{q,u}$ (defined in Section \ref{subsec:ECSI}), consisting of $g_{q,k,u}^\text{eff}$ for all $k \in z_\ell$. Note that we normalize the precoding to decouple DBF from PD optimization discussed later. In the presence of DBF, the ECSI for each UE should be updated so as to also capture the impact of DBF. To this end, we define
\begin{align}
h_{q,k,u}(z_\ell)={\mathbf{g}^{\text{eff}}_{q,u}}^T (z_\ell) \mathbf{d}_{q,k}(z_\ell), \quad k \in z_\ell, \label{eq:effChan}
\end{align}
where, $[\mathbf{d}_{q,u_1}(z_\ell),...,\mathbf{d}_{q,u_{|z_\ell|}}(z_\ell)] \triangleq \mathbf{F}_q^{\text{DBF}}(z_\ell)$, and
\begin{align}
\mathbf{F}_q^{\text{DBF}}(z_\ell) = 
    \begin{cases}
    \mathbf{F}_q^{\text{ZF}}(z_\ell), \text{if ZF-DBF}\\
    \mathbf{I}(|z_\ell|), \text{if N-DBF}\
    \end{cases},
\end{align}
where $\mathbf{I}(m)$ is the identity matrix of size $m$ and $h_{q,k,u}(z_\ell)$ is the end-to-end effective channel after applying both ABF and DBF. It will be used in SINR expressions in Sec.~\ref{sec:offline}. 

\vspace{-10pt}
\subsection{Rate Function} \label{subsec:ratefunc}
The rate function $f(\cdot)$ translates the SINR of a UE in a given RBL into a data rate in bps. We consider the rate function derived from practical MCSs instead of the widely used Shannon capacity formula. More specifically, we consider the data rate to be a piecewise constant function of SINR made of $L$ levels corresponding to the $L$ MCSs:
\begin{equation}
  f(\gamma)=B_r\big(s_1\mathbbm{1}_{[\Gamma_1,\Gamma_2)}(\gamma) + 
  \cdots + s_{L} \mathbbm{1}_{[\Gamma_{L},\infty)}(\gamma)\big), \label{eq:true_rate}
\end{equation} 
where $B_r$ is the size in Hz of an RBL, $\mathbbm{1}_A(x)$ denotes the indicator function with a value of 1 if $x \in A$ and zero otherwise, $\Gamma_l$ is the SINR decoding threshold for MCS $l$ and $s_l$ is the spectral efficiency of MCS $l$ measured in bps/Hz. 
As it is difficult to deal with such a rate function in optimization problems due to discontinuities, we introduce a convex approximation of $f(\cdot)$ which will be used later, namely: 
\begin{align} \label{xxx}
    \tilde{f}(\gamma)=B_r \Big(\min\big\{a \log \big(1+\gamma/b\big), s_{\max} \big\}\Big),
\end{align}
where $a$ and $b$ are parameters controlling the approximation and $s_{\max}=s_L$ is the spectrum efficiency of the highest MCS level.
Fig.~\ref{fMCS} shows the piece-wise constant rate function, a convex tight approximation $\tilde{f}_1(\gamma)$ corresponding to $(a,b)= (1.237,1.952)$ and an upper-bound approximation $\tilde{f}_2(\gamma)$ corresponding to $(a,b)= (1.181,1.253)$, where the set of MCSs are adopted from 3GPP \cite[Table 7.2.3-1]{3gpp-rate}. We will use these functions in our evaluations. 
\vspace{-5pt}
\section{Planning (Offline RRM Study)} \label{sec:offline}
In this section, we provide an offline planning RRM study where beam selection, user selection,  PD, and optionally DBF, are jointly optimized. While the solutions to the RRM optimization problems can be of high complexity, the evaluation results provide engineering insights on the impact of various system parameters as well as the importance of different procedures such as DBF and PD. Further, the solutions will provide performance targets for designing online low-complexity RRM schemes which will be discussed in Section~\ref{sec:online}.

\vspace{-10pt}
\subsection{Fairness}


Fairness is crucial in  RRM studies. In this paper, we consider proportional fairness \cite{PF1,PF2}, where the objective is to maximize the product of the UE throughputs over a certain horizon. To elaborate, keeping in mind that in the online study, we will work in a myopic way,   having kept track of the past, we formulate offline problems that maximize, at the beginning of MB $i$, the product of the UEs' throughputs over the past (and the present) $W+1$ MBs, i.e., 
\begin{align}\label{OF}
    & \max \prod_{u \in \mathcal{U}}  WR_u(i)+\lambda_u(I) \equiv \max \sum_{u \in \mathcal{U}}  \log(WR_u(i)+\lambda_u(I)) \nonumber \\ 
    & \equiv \max \sum_{u \in \mathcal{U}}  \log(1+ \frac{\lambda_u(i)}{WR_u(i)}), 
\end{align}
where $\lambda_u(i)$ is the throughput of UE $u$ within MB $i$ and $R_u(i)$ is the average throughput received by $u$ over the past $W$ MBs. We assume that a BA period is much larger in time than $W$ MBs. Additionally, at the end of MB $i$, we use a moving average of size $W$ to update $R_u(i)$ as $R_u(i+1)=(1-1/W) R_u(i)+\lambda_u(i)/W, \forall u \in \mathcal{U}$. We highlight that since the effective channels are estimated in parallel for all the RBLs within each MB, an MB  is the largest time-frequency resource set over which we can optimize RRM.  For brevity of notation, we will remove the MB index $i$. Following the PF literature, we approximate the objective function in \eqref{OF} by $\sum_{u\in \mathcal{U}} \frac{\lambda_u}{W R_u}$
assuming $WR_u \gg \lambda_u$ and using the approximation $\log(1+x) \approx x$ for $x \ll 1$.

\vspace{-10pt}
\subsection{Problem Formulation}
We formulate a generic RRM optimization problem. We will specialize it later depending if DBF is included or not and if PD is optimized or not.
The problem is formulated for a given MB, keeping track of the $R_u$'s. 
At each time slot, at most $L=\min(K,|\mathcal{B}_p|)$ beams can be selected (recall that $\mathcal{B}_p$ is the set of all BS beams that are preferred by at least one user) and the selected beam set should be kept the same for all PRBs in that time slot. We construct \emph{beforehand} super set $\mathcal{L}$ which consists of all the possible beam sets $\ell$ of $L$ preferred beams, 
i.e., we define
\begin{align}
    \mathcal{L}=\{\ell:\ell=\{b_1^\ell,...,b_j^\ell,...,b_{L}^\ell\},\ b_j^\ell \in \mathcal{B}_p,\ 1 \leq j \leq L\}.
\end{align}

Given that a beam set $\ell$ is selected for a time slot, the BS will select  UE sets. Note that these UE sets could be of size less than or equal to $L$. A UE can only be in a selected UE set, if its preferred beam is in the selected beam set $\ell$, and no UEs in a UE set should share the same preferred beam. Note that it is possible for multiple UEs to prefer the same beam. Therefore, a beam set $\ell$ can be mapped to multiple UE sets.  We construct \emph{beforehand} the set of all distinct possible UE sets corresponding to the beam set $\ell$  defined as:
\begin{align}
    \mathcal{M}_\ell = & \big\{z_\ell:z_\ell=\{u_1,...,u_j,...,u_{|\ell'|}\}, \nonumber \\
    & \ u_j \in \mathcal{U}, b(u_j)=b_j^{\ell'},\ell'=\{b_1^{\ell'},...,b_{|\ell'|}^{\ell'}\} \subset \ell \big\},
\end{align}
where $b(u_j)$ is the index of the preferred ABF vector for UE $u_j$, and $\ell'$ is a subset of $\ell$. Note that we can compute $\mathcal{L}$ and $\mathcal{M}_\ell$ for each $\ell$ after BA and before RRM. Specifically, given a beam set $\ell$, we choose at most one UE from the set of UEs preferring beam $b \in \ell$  when constructing a UE set $z_\ell$.
 
Next, we introduce the variables of the joint RRM optimization problem. We first define $\alpha_\ell$ as the fraction  of time slots occupied by beam set $\ell$ in an MB. This variable does not change with $q$ as the selected beams need to remain the same within a time slot. Moreover, as the channel coefficients do not change within a RBL, it does not make any difference where in time within an MB a beam set $\ell$ is selected, as long as the fraction $\alpha_\ell$ is preserved. We have 
\vspace{-5pt}
\begin{align}
&\sum_{\ell \in \mathcal{L}} \alpha_\ell =1, 
~~~0 \leq \alpha_\ell \leq 1,~\forall \ell\in\mathcal{L}, \label{C1}\\
&N_T\alpha_\ell \in \mathbb{Z},~
\forall \ell\in\mathcal{L}. \label{Z1}
\end{align}

\vspace{-5pt}

The last constraint is because we have $N_T$ time slots within each MB and the number of time slots assigned to a beam set should be an integer.

Given the BS selected beam set $\ell$, various UE sets can be scheduled in each of the $QN_F$ vertical PRBs of each time slot in an MB. Let $\beta_\ell^{q}(z_\ell)$ be the fraction of PRBs within RBL $q$ that UE set $z_{\ell}$ is selected given that $\ell$ is active. As the channel coefficients are the same in each PRB of an RBL, it does not make any difference which PRBs are allocated to which UE set as long as the fractions are preserved. We have 
\begin{align}
& \sum_{z_\ell \in \mathcal{M}_\ell} \beta^q_\ell (z_\ell) \leq 1, \beta^q_\ell (z_\ell) \geq 0,  ~\forall \ell, q, z_\ell \label{C2}\\
& N_F \beta^q_\ell(z_\ell) \in \mathbb{Z}, ~\forall \ell, q,  z_\ell, \label{Z2}
\end{align}

\vspace{-5pt}

where $\mathcal{Q} = \{1,2,,\ldots,Q\}$ is the set of vertical RBL indices within an MB. Equation \eqref{Z2} is due to the fact that we have $N_F$ PRBs at each time slot within each RBL and the number of PRBs assigned to a UE set should be integer.


Note that the computation of $\alpha_\ell$ and $\beta^q_\ell(z_\ell)$ will determine the allocation of the PRBs in an MB to different UE sets.
 
Next, we introduce the variables related to PD. Recall that we assume uniform power allocation over different frequency subchannels, i.e., we assume that the BS power budget for each PRB is $\frac{P_\text{BS}}{Q N_F}$, where $P_\text{BS}$ is the total BS power budget. Let $p_\ell^{q,u} (z_\ell)$ be the power allocated to UE $u$ in UE set $z_\ell$ within the PRBs occupied by $z_\ell$ inside RBL $q$. Note that we have assigned the same per PRB power for all PRBs allocated to a set $z_\ell$ in RBL $q$. In other words, we consider optimizing PD  with RBL-level granularity. While one can consider finer granularities (such as PRB-level) for the optimization of PD variables, it would significantly increase the complexity of the problem making it difficult to solve even in an offline manner. 
 
In the design of ABF and DBF (if present), we ensure the norm of the overall HBF vector for each UE within a UE set is 1, i.e., $ \lVert \mathbf{F}^\text{ABF}(z_\ell)\mathbf{d}_{q,u}(z_\ell) \lVert_F = 1$ for all $u \in z_\ell$, where $\mathbf{F}^\text{ABF}(z_\ell)$ and $\mathbf{d}_{q,u}(z_\ell)$ (defined in Section \ref{subsec:DBF}) correspond to ABF and DBF, respectively. This assumption enables us to decouple the PD from BF.  



For a PRB in RBL $q$ allocated to UE set $z_\ell$ we have

\vspace{-5pt}

\begin{align}
    & \sum_{u \in z_\ell} p_{\ell}^{q,u}(z_\ell) \leq \frac{P_\text{BS}}{Q N_F} \;\;\; \forall \ell \in \mathcal{L},  q \in \mathcal{Q},  z_\ell \in \mathcal{M}_\ell \label{C5}\\
    & 0 \leq p_{\ell}^{q,u}(z_\ell), \forall \ell \in \mathcal{L}, z_\ell \in \mathcal{M}_\ell, u \in z_\ell, q \in \mathcal{Q}   \label{C6}
\end{align}

The SINR of UE $u$ in a PRB within RBL $q$ allocated to UE set $z_\ell$ is

\vspace{-5pt}

\begin{align}
    & \gamma_\ell^{q,u}(z_\ell)=\left\{
    \begin{array}{l}
    \frac{|h_{q,u,u}(z_\ell)|^2 p_{\ell}^{q,u}(z_\ell)}{
    \sum\limits_{\begin{subarray}{c} n\in z_\ell\\ n \neq u \end{subarray}} {|h_{q,n,u}(z_\ell)|^2 p_{\ell}^{q,n}(z_\ell)} + \sigma_\text{PRB}^2}, \text{if } u \in z_\ell\\
    0, \text{otherwise }\\
    \end{array}
    \right. \label{C7}
\end{align}
where $\sigma_\text{PRB}^2$ is the noise power on a PRB. Note that SINR $\gamma_\ell^{q,u}(z_\ell)$ includes $h_{q,i,u}(z_\ell), \forall i \in z_\ell$ which can be computed using \eqref{eq:effChan}. The DL data rate of UE $u$ in a PRB within RBL $q$ allocated to UE set $z_\ell$ is defined as
\begin{align}
r_\ell^{q,u}(z_\ell) =  f\big(\gamma_\ell^{q,u}(z_\ell)\big), \forall u,  q,  z_\ell, \label{C8}
\end{align}
where $f(.)$ is the rate function described in Section \ref{subsec:ratefunc}. Given $\alpha_\ell$, $\beta^q_\ell(z_\ell)$ and $r_\ell^{q,u}(z_\ell)$, we can compute the throughput $\lambda_u$ of  UE  $u$ in the MB as
\begin{align}
    \lambda_u = \sum_{\ell \in \mathcal{L}} \alpha_\ell \sum_{z_\ell \in \mathcal{M}_\ell}\sum_{q \in \mathcal{Q}} \beta^q_\ell (z_\ell) r_\ell^{q,u}(z_\ell), \quad \forall u \in \mathcal{U}. \label{C9}
\end{align}
Next, we formulate the general RRM optimization problem for an MB. Given $\mathbf{g}^\text{eff}_{q,u}$, $\mathbf{F}_q^{\text{DBF}}$,$\mathbf{F}_q^{\text{ABF}}$, $\mathcal{U}$, $\mathcal{Q}$, $\mathcal{L}$, $\mathcal{M}_\ell$, $R_u$, $W$, $N_T$ and $N_F$, we formulate
\begin{align*}
    \mathbf{\Pi}: \max_{\beta^q_\ell(z_\ell), \alpha_\ell, p_{\ell}^{q,u}(z_\ell), \gamma_\ell^{q,u}(z_\ell), r_\ell^{q,u}(z_\ell), \lambda_u} \sum_{u\in \mathcal{U}} \frac{\lambda_u}{W R_u}, \nonumber \\
    ~~ \text{s.t. (\ref{C1}), (\ref{Z1}), (\ref{C2}), (\ref{Z2}), (\ref{C5}), (\ref{C6}), (\ref{C7}), (\ref{C8}), (\ref{C9})}.
\end{align*}
Problem $ \mathbf{\Pi}$ is non-convex, non-linear and includes integer constraints. To obtain solutions, we first relax it by removing integer constraints in \eqref{Z1} and \eqref{Z2}. Further, we introduce the following change of variables $\xi^q_\ell (z_\ell) \triangleq \alpha_\ell \beta^q_\ell (z_\ell)$ to eliminate the product of variables in (\ref{C9}). Subsequently, we formulate the relaxed transformed problem as
\vspace{-10pt}
\begin{align}
    \mathbf{\Pi}^\text{R}: & \max_{\xi^q_\ell(z_\ell), \alpha_\ell, p_{\ell}^{q,u}(z_\ell), \gamma_\ell^{q,u}(z_\ell), r_\ell^{q,u}(z_\ell), \lambda_u} \sum_{u\in \mathcal{U}} \frac{\lambda_u}{W R_u} \nonumber\\
\noindent \text{s.t. }
    & (\ref{C1}), (\ref{C5}), (\ref{C6}), (\ref{C7}), (\ref{C8}), \nonumber\\
    & \sum_{z_\ell \in \mathcal{M}_\ell} \xi^q_\ell (z_\ell) \leq \alpha_\ell, \;\;\; \forall \ell \in \mathcal{L}, \forall q \in \mathcal{Q}, \label{C3a}\\
    & 0 \leq \xi^q_\ell (z_\ell) \leq 1,  \;\;\;\forall \ell \in \mathcal{L}, \forall q \in \mathcal{Q}, \forall z_\ell \in \mathcal{M}_\ell, \label{C4a}\\
    & \lambda_u = \sum_{\ell \in \mathcal{L}} \sum_{z_\ell \in \mathcal{M}_\ell}\sum_{q \in \mathcal{Q}} \xi^q_\ell (z_\ell) r_\ell^{q,u}(z_\ell), \;\;\; \forall u \in \mathcal{U}. \label{C9a}
\end{align}
The variables are all continuous in Problem $\mathbf{\Pi}^\text{R}$. Note that the optimal objective of Problem $\mathbf{\Pi}^\text{R}$ is an upper-bound for that of Problem $\mathbf{\Pi}$. Next, we provide methods to solve Problem~$\mathbf{\Pi}^\text{R}$ for three different cases: (i) EPD (equal power distribution) with or without DBF, (ii) OPD without DBF, and (iii) OPD with ZF-DBF.
\vspace{-10pt}
\subsection{Solution Method for EPD}
With equal power distribution,  we have $p_{\ell}^{q,u}(z_\ell)=P_\text{BS} /(|z_\ell|Q N_F)$. Consequently, the SINRs defined in \eqref{C7} and the data rates introduced in \eqref{C8} can be computed prior to RRM optimization. Note that we do not set $p_{\ell}^{q,u}(z_\ell)=P_\text{BS} /(LQ N_F)$ to allow for the use of UE sets with smaller cardinality if it is beneficial. Using EPD, the RRM problem can be formulated as follows. Given  $\mathbf{g}^\text{eff}_{q,u}$, $\mathbf{F}_q^{\text{DBF}}$,$\mathbf{F}_q^{\text{ABF}}$, $\mathcal{U}$, $\mathcal{Q}$, $\mathcal{L}$, $\mathcal{M}_\ell$, $R_u$, $W$, $N_T$,  $N_F$ and $r_\ell^{q,u}(z_\ell)$:
\begin{align*}
    \mathbf{\Pi}_\text{EPD}:& \max_{\xi^q_\ell(z_\ell), \alpha_\ell, \lambda_u}  \sum_{u\in \mathcal{U}} \frac{\lambda_u}{WR_u}, 
\quad \text{s.t. }
     (\ref{C1}), (\ref{C2}), (\ref{C3a}), (\ref{C4a}), (\ref{C9a}).
\end{align*}
Problem~$\mathbf{\Pi}_\text{EPD}$ is a linear program (LP) which can be optimally solved using a commercial solver. Note that the existence or not  of DBF does not impact the solution method. The reason is that having DBF only impacts the SINRs (see \eqref{C7}) which are computed prior to RRM given EPD. In summary, in the EPD case, we can obtain the optimal solution for the RRM problem with or without DBF. 
\vspace{-10pt}
\subsection{Solution Methods for OPD}
\label{subsec:OPD}
While EPD simplifies the optimization, it restricts the performance. To further improve the performance, we consider the OPD case where the $p_{\ell}^{q,u}(z_\ell)$'s are kept as optimization variables. In this case, we need to solve Problem~$\mathbf{\Pi}^\text{R}$ which can be reformulated as:
\begin{align}
\mathbf{\Pi}_\text{OPD}: &\max_{\xi^q_\ell(z_\ell), \alpha_\ell, p_{\ell}^{q,u}(z_\ell), \gamma_\ell^{q,u}(z_\ell) r_\ell^{q,u}(z_\ell), \lambda}  \lambda, \nonumber\\
\text{s.t. }
    &(\ref{C1}), (\ref{C2}), (\ref{C5}), (\ref{C6}), (\ref{C7}), (\ref{C8}),(\ref{C3a}), (\ref{C4a}),   \nonumber\\
    &\lambda = \frac{1}{W} \sum_{\ell \in \mathcal{L}} \sum_{z_\ell \in \mathcal{M}_\ell}\sum_{q \in \mathcal{Q}} \xi^q_\ell (z_\ell) \sum_{u\in \mathcal{U}} \frac{r_\ell^{q,u}(z_\ell)}{R_u}. \nonumber 
\end{align}
{Due to the independence of $\xi^q_\ell (z_\ell)$ and $r_\ell^{q,u}(z_\ell)$, Problem $\mathbf{\Pi}_\text{OPD}$ can be decoupled into the two sub-problems discussed next. The first sub-problem is a weighted sum-rate optimization which optimizes the PD variables given $\ell$, $z_\ell$ and $q$:
\begin{align*}
 \noindent \mathbf{\Pi}^\text{P1}_\text{OPD}(\ell,z_\ell,q):  & \max_{p_{\ell}^{q,u}(z_\ell), \gamma_\ell^{q,u}(z_\ell), r_\ell^{q,u}(z_\ell)} \sum_{u\in \mathcal{U}} \frac{r_\ell^{q,u}(z_\ell)}{R_u}, \nonumber \\
\text{s.t. } &
 (\ref{C5}), (\ref{C6}), (\ref{C7}), (\ref{C8}) \nonumber
\end{align*}
We will provide solution methods to Problem $\mathbf{\Pi}^\text{P1}_\text{OPD}(\ell,z_\ell,q)$ for scenarios with and without DBF-ZF below. 
Let $\Gamma_\ell^q(z_\ell)$ denote the optimal objective value of Problem~$\mathbf{\Pi}^\text{P1}_\text{OPD}(\ell,z_\ell,q)$.
Given $\Gamma_\ell^q(z_\ell)$ for all $q \in \mathcal{Q}$, $\ell \in \mathcal{L}$ and $z_\ell \in \mathcal{M}_\ell$, we can then solve the second sub-problem as follows:
\begin{align*}
 \noindent \mathbf{\Pi}^\text{P2}_\text{OPD}:     & \max_{\xi^q_\ell(z_\ell), \alpha_\ell, \lambda}   \frac{1}{W} \sum_{\ell \in \mathcal{L}} \sum_{z_\ell \in \mathcal{M}_\ell}\sum_{q \in \mathcal{Q}} \xi^q_\ell (z_\ell) \Gamma_\ell^q(z_\ell), \nonumber \\
\text{s.t. } &
     (\ref{C1}), (\ref{C2}), (\ref{C3a}), (\ref{C4a})  
\end{align*}
which is an LP that can be optimally solved by efficient solvers. Consequently, to solve problem $\mathbf{\Pi}_\text{OPD}$ it is sufficient to solve sub-problem~$\mathbf{\Pi}^\text{P1}_\text{OPD}$ for all $\ell$, $q$ and $z_\ell$ to obtain the values $\Gamma_\ell^q(z_\ell)$ and then use them to solve sub-problem~$\mathbf{\Pi}^\text{P2}_\text{OPD}$. Note that no iteration is required between the sub-problems but an exhaustive computation of $\Gamma_\ell^q(z_\ell)$ for all $\ell$, $q$ and $z_\ell$ is.

Next, we provide solution methods for Problem $\mathbf{\Pi}^\text{P1}_\text{OPD}(\ell,z_\ell,q)$ for the scenarios with and without ZF-DBF. 
\subsubsection{Solving \texorpdfstring{$\mathbf{\Pi}^\text{P1}_\text{OPD}(\ell,z_\ell,q)$}{Lg} for the scenario with ZF-DBF} \label{subsection:OPD-ZF-DBF}
ZF-DBF eliminates the inter-beam interference, i.e., the residual interference after BA. Therefore, with ZF-DBF we have $h_{q,n,u}(z_\ell)=0$ if $n \neq u$ and the  SNR (it is now an SNR since there is no interference) expression \eqref{C7} can be simplified as
\begin{align}
    \gamma_\ell^{q,u}(z_\ell)= & \left\{
    \begin{array}{l}
    \frac{|h_{q,u,u}(z_\ell)|^2 p_{\ell}^{q,u}(z_\ell)}{\sigma_\text{PRB}^2}, \text{ if } u \in z_\ell\\
    0, \text{ otherwise }\\
    \end{array},
    \right. \label{C7p}
\end{align}


In that case, the only non-convex part in $\mathbf{\Pi}^\text{P1}_\text{OPD}(\ell,z_\ell,q)$ is Constraint~\eqref{C8}, since it involves the piece-wise constant rate function $f$ (modeling MCSs). To convexify the problem, we approximate $f(\gamma)$ by the convex rate function $\tilde{f}_1(\gamma)$ (resp. $\tilde{f}_2(\gamma)$) introduced in Section~\ref{subsec:ratefunc} to acquire the feasible (resp. upper-bound) solution and objective value for $\mathbf{\Pi}^\text{P1}_\text{OPD}(\ell,z_\ell,q)$. With the feasible (resp. upper-bound) solution and objective value for $\mathbf{\Pi}^\text{P1}_\text{OPD}(\ell,z_\ell,q)$  for all $\ell$, $q$ and $z_\ell$,  we can solve $\mathbf{\Pi}^\text{P2}_\text{OPD}$ to obtain a feasible (resp. an upper-bound) solution and objective value for $\mathbf{\Pi}_\text{OPD}$. 


Specifically, to obtain the feasible (resp. upper-bound) solution and objective value of $\mathbf{\Pi}_\text{OPD}$, we solve $\mathbf{\Pi}^\text{P1}_\text{OPD}(\ell,z_\ell,q)$ for all $q \in \mathcal{Q}$, $\ell \in \mathcal{L}$ and $z_\ell \in \mathcal{M}_\ell$ after replacing $f(\gamma)$ by $\tilde{f}_1(\gamma)$ (resp. $\tilde{f}_2(\gamma)$). Then, for the feasible solution,  we use the solutions, i.e., the power variables, to calculate the data rate according to the piece-wise constant  rate function $f$, and use the resulting $\Gamma_\ell^q(z_\ell)$ into $\mathbf{\Pi}^\text{P2}_\text{OPD}$ to solve it. 

Our numerical evaluations provided in Section~\ref{subsec:offline-sym} reveal that the feasible solution and the upper-bound are close and hence the optimal solution to our original problem is well approximated by the feasible solution. 

\subsubsection{Solving \texorpdfstring{$\mathbf{\Pi}^\text{P1}_\text{OPD}(\ell,z_\ell,q)$}{Lg} for the N-DBF scenario} \label{subsection:OPD-N-DBF}
In that case, there is inter-beam interference and therefore, the SINR given in Eq.~\eqref{C7} is a non-convex function of the PD variables causing the problem to be highly non-convex. One can only obtain a locally optimal solution by applying a local search around an initial solution using commercial solvers such as MINOS. This implies that the quality of the solution depends on the initial solution for the power values. We have tried various methods for finding an initial solution including (i) setting all the power variables to zero, (ii) using the solution of EPD, and (iii) using the solution of Problem~$\widehat{\mathbf{\Pi}}^\text{P1}_\text{OPD}(\ell,z_\ell,q)$ as the initial solution. Our numerical evaluations revealed that using method (iii) consistently provides a better performance. Given the initial point, using an approach similar to the one used in Section~\ref{subsection:OPD-ZF-DBF}, we transform the problem using a tight convex approximation of the rate function $f$ and find a local optimum using MINOS in AMPL. 

\vspace{-10pt}

\subsection{Extension to Multiple Pairs of Preferred Beams Per UE} \label{MultipeBeamPair}


Until now, we have assumed that one beam pair is selected for each UE during the BA process. Having more than one pair of preferred beams per UE increases the flexibility of beam set and UE set selections which can potentially improve the performance (at a cost of higher BA and CE overhead of course). In this subsection, we investigate the RRM problem when multiple (say $M$) beam pairs are selected per UE during the BA process. Note that the UE is still assumed to have a single RF chain. Therefore, at most one of the beam pairs can be selected at a time for each UE. To elaborate on the setting, let $(\mathbf{w}_{u,i}^*,\mathbf{v}_{u,i}^*)$, $i\in\{1,\ldots,M\}$ denote the $i$-th beam pair chosen for UE $u$ during the BA process. We define $\mathcal{C}_b^\star= \{\mathbf{w}_{u,i}^*\}_{u \in \mathcal{U},i \in \{1,\ldots, M\}}$ as the set of all distinct preferred ABF vectors at the BS. Also, we define $\mathcal{B}_p=\{b_1,...,b_{|\mathcal{C}_b^\star|}\}$ as the set of indices corresponding to the ABF vectors in $\mathcal{C}_b^\star$. 

During CE,  the effective CSI of each UE for all its $m$ pairs of preferred beams is estimated. Let $g_{q,u,u,i,i}^\text{eff} \in \mathbb{C}$, $i\in\{1,\ldots,M\}$ be the effective channel between the BS and UE $u$ in any PRB of RBL $q$, when using its $i$-th preferred beam pair. Furthermore, let $g_{q,n,u,j,i}^\text{eff} \in \mathbb{C}$, $i,j\in\{1,\ldots,M\}$ denote the effective interference channel in any PRB of RBL $q$ between the BS and UE $u$, when UE $u$ applies its $i$-th preferred UE beam and the BS applies the $j$-th preferred BS beam of UE $n$. We have $g_{q,n,u,j,i}^\text{eff}=\left({\mathbf{v}^*_{u,i}}\right)^H \mathbf{G}_{q,u}\left(\mathbf{w}_{n,j}^*\right)$. With beam set $\ell$, we can generate $\mathcal{M}_\ell$, the set of all possible UE sets $z_\ell$ corresponding to beam set $\ell$. Note that there is a special case when more than one of the preferred BS beams of a UE belong to $\ell$. In such cases, a UE should not be selected multiple times in $z_\ell$. Hence, we remove instances of $z_\ell$ that include a UE more than once from $\mathcal{M}_\ell$ since we assume there is at most one stream per UE. Given beam set $\ell$ and UE set $z_\ell$, the indices of the preferred beams, i.e., $i$ and $j$ in $g_{q,n,u,j,i}^\text{eff}$ are known for the UEs in $z_\ell$. Let $z_\ell = \{u_1,u_2,\ldots,u_{|z_\ell|}\}$. For every $u\in z_\ell$, we define $\mathbf{g}^\text{eff}_{q,u}(z_\ell) = [g_{q,u_1,u,j_{u_1},i_u}^\text{eff},...,g_{q,u_{|z_\ell|},u,j_{u_{|z_\ell|}},i_u}^\text{eff}]^T \in \mathbb{C}^{|z_\ell| \times 1}$, where $j_{u_n}\in{1,\ldots,M}$ is the index of the corresponding preferred BS beam for UE $u_n$ given beam set $\ell$. Similarly, $i_u\in\{1,\ldots,M\}$ is the index of the corresponding preferred UE beam for UE $u$ given beam set $\ell$. Subsequently, we can proceed with defining DBF based on $\mathbf{g}^\text{eff}_{q,u}(z_\ell)$ as described in Sec.~\ref{subsec:DBF}.

The problem formulations and solutions provided in the previous subsections of this section are  applicable to the case of multiple pairs of preferred beams per UE. 
We will provide simulation results in Sec.~\ref{subsec:offline-sym} to evaluate the impact of increasing the number of preferred beam pairs per UE to two.

\vspace{-10pt}

\subsection{Numerical Evaluations for Offline RRM Study} \label{subsec:offline-sym}

In this section, we present the numerical results of our offline study. The results enable us to evaluate the impact of PD, ZF-DBF as well as system parameters such as $K$, $B_b$, and $B_u$. The results also demonstrate the impact of beam selection constraint ConB justifying the need for performing RRM optimization jointly for an MB as opposed to a per-subchannel optimization approach proposed in the literature.

We consider the DL of a small cell of radius $r=75$ m within which the UEs are distributed uniformly. The BS height is set to 10 m and the BS power budget is considered to be $P_{BS} = 27$~dBm. We assume that BS and UEs are equipped with uniform linear antenna arrays (ULAs) with half-wavelength inter-element spacing. The system operates at $28$~GHz and the bandwidth is assumed to be $100$~MHz. Furthermore, regarding the size of the report blocks, we assume $N_F=6$,  leading to $B_r = 4.32$~MHz and $Q=22$. We have chosen the value of $N_F$ based on the channel model introduced later such that the channel coefficients can be considered constant within each RBL. To elaborate, we have generated $5\times 10^4$ \emph{channel realizations} by generating 50 realizations with 10 users and 100 MBs and calculated the root mean squared (RMS) delay for each realization. We have chosen $N_F$ to ensure that $\frac{1}{\bar{\tau}_\text{rms}} \gg B_r$ 
where ${\bar{\tau}_\text{rms}}$ is the mean of the RMS delay over different realizations. 
Additionally, we assume $N_T=20$, leading to $T_r=5$~msec which is typical for low-mobility scenarios in mmWave frequency bands \cite{Du2017}. We adopt the set of discrete MCSs from \cite[Table 7.2.3-1]{3gpp-rate}. We use the method in \cite{CODEBOOK1} to generate the ABF codebooks (necessary for BA) for BS and UE, where we set the optimization parameters $M$ and $T$ as 15 and 3, respectively. We refer the interested reader to \cite{CODEBOOK1} for further details. The rest of the simulation parameters are listed in Table \ref{Param}. During BA, we assume that the beam pairs are chosen based on the SNR they provide, i.e. the beam pairs leading to the highest SNRs are selected. Additionally, we neglect the overhead of BA and CE and the processing time required for RRM since this is an offline study. Furthermore, we assume there are frequency guard-bands to prevent inter-band interference and neglect the corresponding overhead.

For our numerical evaluations, we adopt a multi-ray wide-band channel model which incorporates multiple scattering clusters, and each cluster comprises a number of spatial paths \cite{Channel3}. The MIMO channel between BS and UE $u$ at RBL $q$ is defined as the following $N_u \times N_b$ matrix:
\vspace{-5pt}
\begin{align}
    \mathbf{G}_{q,u} = \frac{1}{\sqrt{N_\text{path}}} \sum_{d=1}^{N_\text{cluster}} \sum_{l=1}^{N_\text{path}} g_{d,l}^{u} e^{-j 2\pi \tau_{d,l}^u f_q} \mathbf{a}_\text{RX}(\phi_{d,l}^u) \mathbf{a}_\text{TX}^\text{H}(\theta_{d,l}^u), \nonumber
\end{align}
where $N_\text{cluster}$ is the number of clusters, $N_\text{path}$ is the number of paths in each cluster, $g_{d,l}^{u}$ is the  coefficient of the path $l$ of cluster $d$, $f_q=f_c-\frac{BW}{2}+(q-1)\frac{BW}{Q}$ is the center frequency of RBL $q$, and $\tau_{d,l}^u={\tau_0}_{d}^u+{\tau_1}_{d,l}^u$ in which ${\tau_0}_{d}^u$ is the group delay of the $d$-th cluster and ${\tau_1}_{d,l}^u$ is the relative path delay of the $l$-th path in the $d$-th cluster. It is assumed that $g_{d,l}^{u}=\sigma_d^u \sqrt{\kappa_{d,l}^{u}} e^{j \varphi_{d,l}^{u}}$ where $\sigma_d^u$ models path-loss and shadowing, $\kappa_{d,l}^{u}$ models the fraction of the signal power carried by the $l$-th path within the $d$-th cluster, and $\varphi_{d,l}^{u} \sim \mathcal{U}(0,2\pi)$ models the phase of the $l$-th path within the $d$-th cluster. We assume $\sigma_d^u=\sqrt{(v_d^u 10^{-0.1PL_u})}$, where $v_d^u$ is the fraction of the signal power carried by the $d$-th cluster and $PL_u$ (in dB) is the impact of path loss and shadowing for UE $u$. Moreover, we assume ${\tau_0}_{d}^u$ and ${\tau_1}_{d,l}^u$ are random variables exponentially distributed with mean of $200$ nsec and $20$ nsec, respectively. Channel measurements indicate that the average power decays exponentially with delay \cite{Channel,Channel2}. To model this, we use the approach described in \cite[Section III-C]{Channel} to obtain $v_d^u$ from ${\tau_0}_{d}^u$ and $\kappa_{d,l}^{u}$ from ${\tau_1}_{d,l}^u$.
Moreover, $\theta_{d,l}^u$ and $\phi_{d,l}^u$ represent the AoD and the AoA corresponding to the $l$-th path of cluster $d$ of the channel which are Gaussian distributed around the cluster central angle as described in \cite{Channel}. 
As mentioned earlier, we consider the system within a BA period that we assume equal to $N_o$ MBs. We also  assume the large-scale fading, AoAs and AoDs remain fixed within this BA period. Additionally, $\mathbf{a}_\text{TX}(.)$ and $\mathbf{a}_\text{RX}(.)$ are the array response vectors at the BS and UE $u$ which depend on the antenna array type. For instance, for a ULA including $N$ antennas with half-wavelength inter-element distance, $\mathbf{a}(.)$ is given by $\mathbf{a}(\phi)=[1,e^{j\pi\sin{\phi}},...,e^{j(N-1)\pi\sin{\phi}}]^T$, where $\phi$ is the AoD if the array is at the transmitter and it is AoA if the array is at the receiver. 

Let $\Omega(U)$ be a network realization obtained by generating $U$ uniformly distributed UE positions in the cell excluding a ring with a radius of 6 m around the BS (where path loss models are not valid). It also includes the channel parameters corresponding to these positions. For a given set of system parameters and a value of $U$, we generate $Z=50$ random realizations. For each realization, we carry out BA, once in the beginning, obtain the preferred beams per user, and perform RRM for all the $N_o = 100$~MBs of that BA period. Since our focus is on PF, the performance for one realization is defined to be the Geometric Mean (GM)  of the UEs' throughputs over $100$ MBs, i.e.,:
\vspace{-5pt}
\begin{align}  
    \text{GM}\big(\Omega(U)\big) = \left(\prod_{u=1}^U \frac{1}{N_o} \sum_{o=1}^{N_o}  \lambda_u^{o}\big(\Omega(U)\big)\right)^\frac{1}{U}
\end{align}
\vspace{-5pt}

To obtain the $\lambda_u^{o}\big(\Omega(U)\big)$, we solve the joint RRM problem sequentially for each MB of the $N_o$ MBs. Finally, we take the arithmetic mean of the GMs over all realizations and compute $\overline{\text{GM}}=\sum_\Omega \text{GM}\big(\Omega(U)\big)/Z$ as the performance measure. Next we provide the simulation results.

\begin{table}
\centering
\caption{Simulation Parameters}
\footnotesize
\begin{tabular}{l l}
\hline
\textbf{Parameter} & \textbf{Value}\\
\hline
Number of antennas at BS $N_b$ & 128\\
\hline
Number of antennas at UE $N_u$ & 16\\
\hline
Noise power spectral density $N_0$ & -174 dBm/Hz \\
\hline
Moving average parameter $W$ & 10 \\
\hline
Initial value of $R_u$ & 2\\
\hline
\end{tabular}
\label{Param}
\vspace{-10pt}
\end{table}


Recall that in the EPD case, Problem $\mathbf{\Pi}^\text{R}$ is equivalent to LP Problem $\mathbf{\Pi}_{EPD}$ which can be solved to optimality. Our evaluations show that, for both ZF-DBF and N-DBF cases in the EPD scenario, the solutions satisfy the integer constraints \eqref{Z1} and \eqref{Z2}. Therefore, the relaxation from $\mathbf{\Pi}$ to $\mathbf{\Pi}^\text{R}$ does not cause optimality loss in the EPD scenario. 

For the OPD cases, as explained in Section~\ref{subsection:OPD-ZF-DBF}, we  evaluate the quality of our proposed feasible solution for the OPD scenario with ZF-DBF by comparing it to an upper-bound. 
Fig.~\ref{r1} illustrates feasible and upper-bound values of $\overline{\text{GM}}$ as a function of the number of BS RF chains $K$ for that scenario. Note that the gap between the feasible solution performance and the upper bound is small confirming that the proposed feasible solution is quasi-optimal. We also note that we cannot evaluate the quality of the feasible solution for the OPD scenario with the N-DBF case due to the lack of performance upper-bound. However, this is not critical since
we will show that the performance of the ZF-DBF case (for which a quasi-optimal solution is found) is significantly better than the case with N-DBF justifying the use of ZF-DBF in practice. 

The following remark highlights an important observation from the evaluations which will be used in the design of online RRM schemes discussed in Sec.~\ref{sec:online}. 
\begin{remark} 
\label{remark}
The optimal/feasible solutions  for all scenarios, select one beam set per MB and one UE set per RBL.
\end{remark}

\begin{figure}
    \centering
    \begin{subfigure}[b]{0.45\textwidth}
         \centering
         \includegraphics[width=60mm]{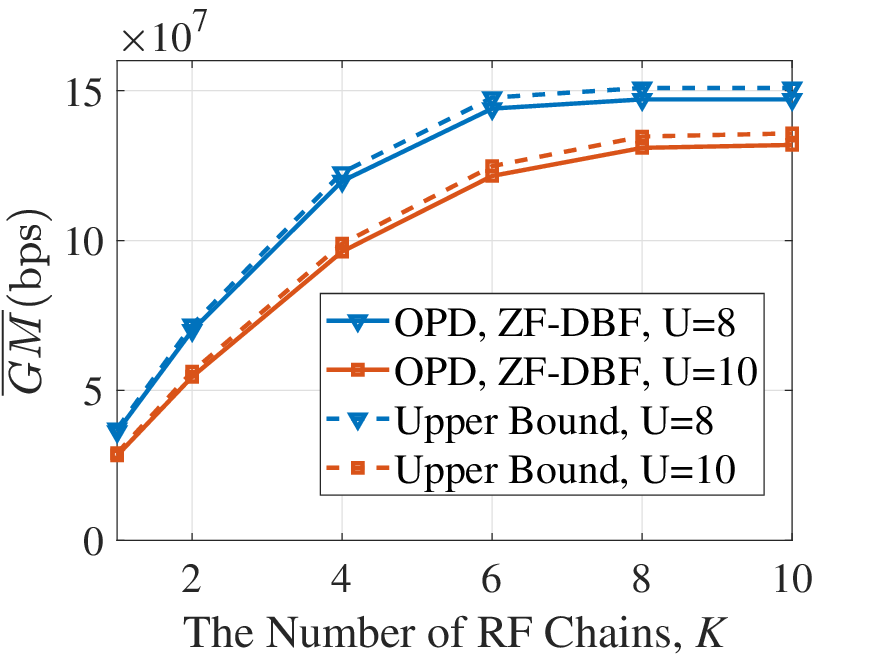}
         \caption{}
         \label{r1}
    \end{subfigure}
    \hfill
    \begin{subfigure}[b]{0.45\textwidth}
         \centering
         \includegraphics[width=60mm]{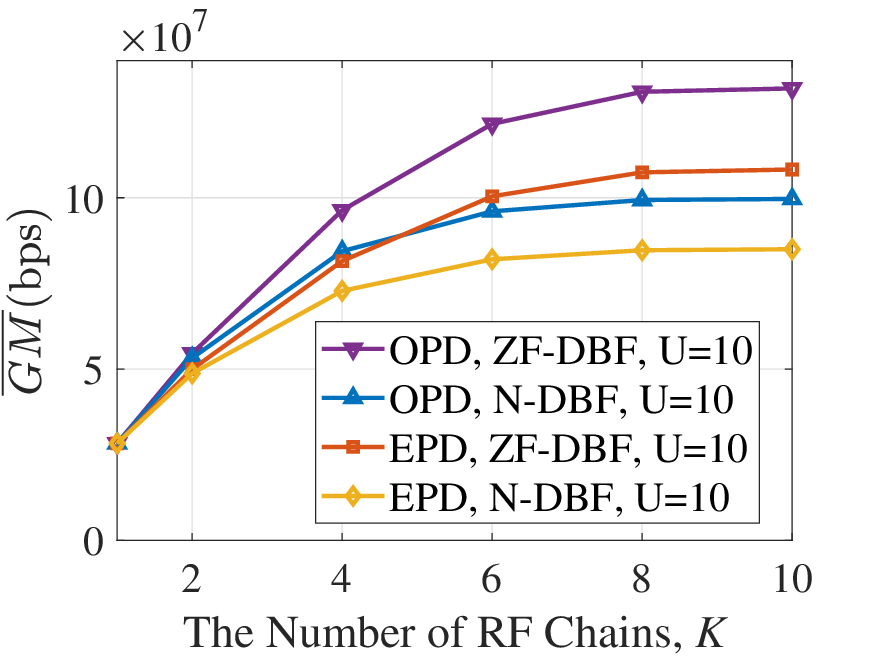}
         \caption{}
         \label{r2}
    \end{subfigure}
    \vspace{-5pt}
    \caption{(a) Quality of feasible solution, and (b) System performance vs. $K$ for different cases EPD/OPD and ZF-DBF/N-DBF. For (a) and (b): $N_b=128$, $N_u=16$, $B_b=32$, $B_u=4$. }
    \vspace{-10pt}
\end{figure}


Next, we evaluate the impact of various RRM procedures such as PD and DBF as well as different system parameters on the system performance. Fig.~\ref{r2} depicts the system performance measure $\overline{\text{GM}}$ as a function of $K$ for different scenarios. The fact that the performance plateaus when $K$ increases above the number of users $U$ is natural since at most one stream can be used per user.
Fig.~\ref{r2} shows the impact of PD and DBF. Considering `EPD, N-DBF' as the baseline scenario, the gain in $\overline{\text{GM}}$ obtained by performing OPD instead of EPD increases as $K$ increases and reaches 17.3\% (w.r.t. the baseline) when $U=10$. We also observe that the gain by replacing N-DBF with ZF-DBF is not large with small $K$, because the number of selected UEs per PRB is small restricting the interference and the gain of ZF-DBF. Increasing $K$ enhances the performance gain of ZF-DBF by up to 27.4\% (w.r.t. the baseline). When both OPD and ZF-DBF are performed, the performance gain (w.r.t. the baseline) reaches 55.2\% for $U=10$. 
The reason for such significant performance gains is the fact that ZF-DBF mitigates the inter-beam interference and OPD improves the fairness among UEs by properly distributing transmit power within the PRBs.

\begin{figure}
    \centering
    \begin{subfigure}[b]{0.45\textwidth}
         \centering
         \includegraphics[width=60mm]{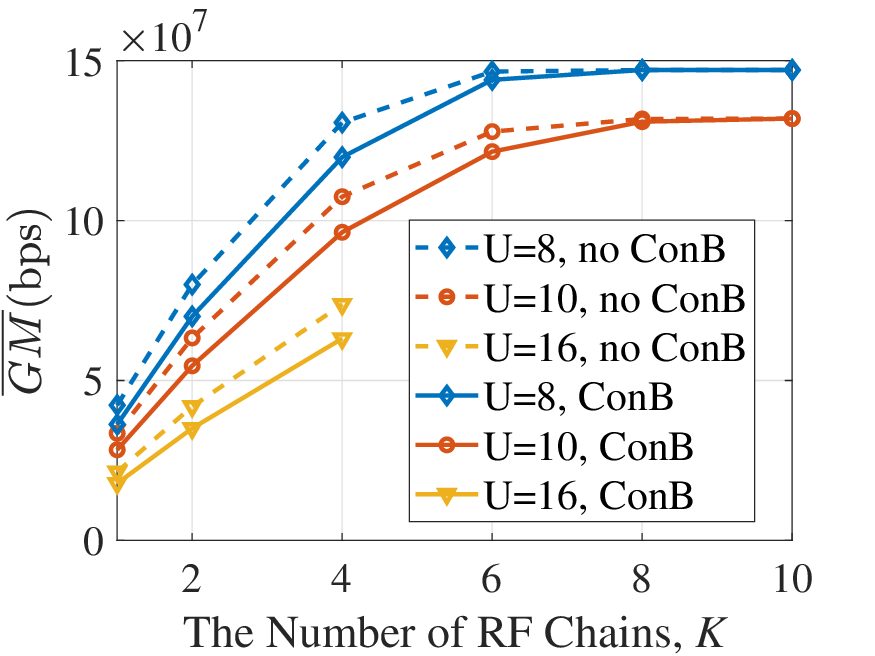}
         \caption{}
         \label{r3}
    \end{subfigure}
    \hfill
    \begin{subfigure}[b]{0.45\textwidth}
         \centering
         \includegraphics[width=60mm]{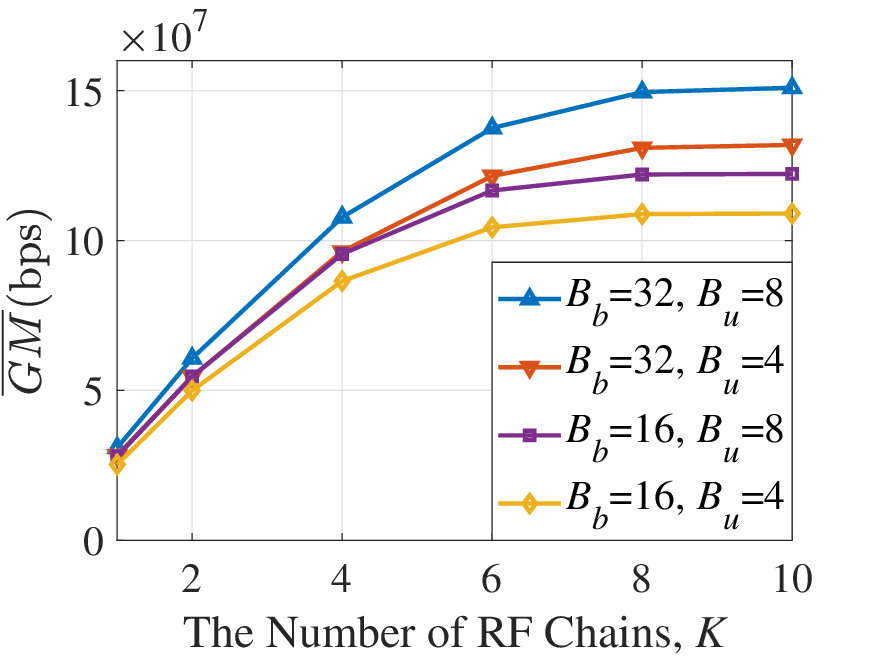}
         \caption{}
         \label{r4}
    \end{subfigure}
    \vspace{-5pt}
    \caption{(a) Impact of ConB on system performance considering OPD, ZF-DBF, $N_b=128$, $N_u=16$, $B_b=32$, $B_u=4$, (b) System performance vs. $K$ for different ABF codebook sizes considering OPD, ZF-DBF, $U=10$, $N_b=128$, and $N_u=16$.}
\vspace{-10pt}
\end{figure}

Fig.~\ref{r3} illustrates the importance of beam selection constraint ConB (i.e., that only one beam set can be selected per time slot). In this figure, we consider the OPD scenario with ZF-DBF and compare the performance when ConB is considered (i.e., performing RRM for an MB with ConB) and when ConB is overlooked (i.e., performing RRM on a per-PRB basis neglecting ConB). 
First, we observe that the performances of `ConB' and `no ConB' are the same if $K \geq U$. This is expected since if $K \geq U$, all the preferred beams can be selected at every time slot which makes beam selection trivial. Hence constraint ConB is automatically satisfied and user selection and PD can be done on a PRB basis.  However, when $K<U$, we observe that overlooking ConB overestimates the performance by as much as 16.5\%, 18.2\% and 20.1\% for $U=8$, $U=10$ and $U=16$, respectively. The over-estimation gets larger as $U$ increases, indicating that the significance of ConB increases with $U$ for the regime where $K<U$. Note that implementation cost and complexity limit the value of $K$ in practice, making $K<U$ a reasonable assumption for practical scenarios. Therefore, it is misleading to overlook ConB in practical scenarios.

Next, we examine the impact of the number of beams $B_b$ (resp. $B_u$) in the BS (resp. UE) ABF codebook on the system performance of the OPD scenario with ZF-DBF where $U=10$ UEs are considered. In Fig.~\ref{r4}, considering the case with $B_b=16$ and $B_u=4$ as the baseline, we observe that doubling $B_b$ can improve the performance by up to 21\%  while doubling $B_u$ can at most provide 12.1\% improvement. Furthermore, Doubling both $B_b$ and $B_u$ can improve the performance by up to 38.5\%. We conclude that increasing the number of beams in the ABF codebooks can considerably boost the system performance. The reason is that increasing $B_b$ and $B_u$ leads to narrower beams which increases the BF gain. Moreover, it is more beneficial to increase the number of beams at BS than at the UEs since increasing $B_b$ not only increases BF gain, but also reduces inter-beam interference due to using narrower beams. We also highlight that BA complexity and overhead increase with increasing $B_b$ and $B_u$ which limits the number of beams we can have during system operation.

\begin{figure}
    \centering
    \begin{subfigure}[b]{0.45\textwidth}
         \centering
         \includegraphics[width=60mm]{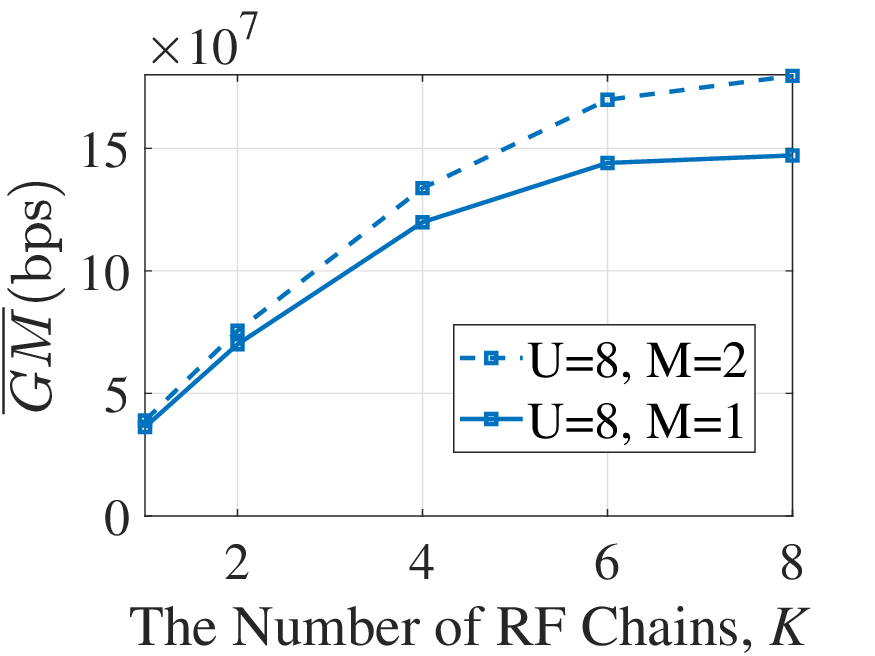}
         \caption{}
         \label{MultipleBeams}
    \end{subfigure}
    \hfill
    \begin{subfigure}[b]{0.45\textwidth}
         \centering
         \includegraphics[width=60mm]{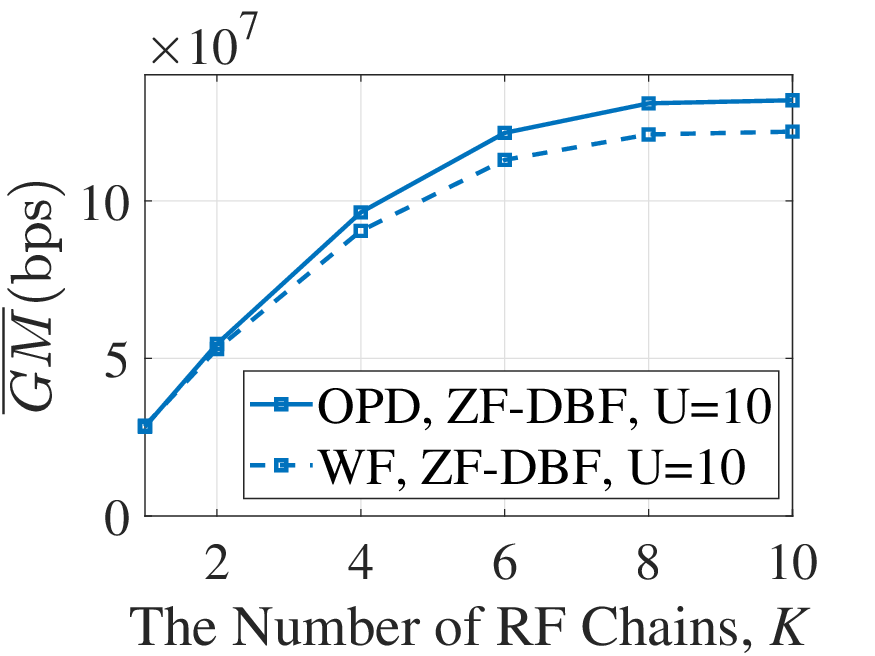}
         \caption{}
         \label{WF}
    \end{subfigure}
    \vspace{-5pt}
    \caption {(a) Impact of the number of preferred beams on system performance considering OPD, ZF-DBF (b) WF performance for ZF-DBF. For (a) and (b): $N_b=128$, $N_u=16$, $B_b=32$, $B_u=4$.}
\vspace{-10pt}
\end{figure}

Fig.~\ref{MultipleBeams} compares the performance of the case with one pair of preferred beams ($M=1$) per UE and the case with two pairs ($M=2$) for each UE after BA. We observe not much gain of performance for $K$ small and up to 22.1\% gain for larger $K$
with $M=2$ due to the more flexibility in beam and user selection which can reduce the interference. However, please note that the cost of such an improvement is a larger amount of ECSI ($M^2$ times) to be measured. 


Water-filling (WF) is a well-known algorithm for power optimization in RRM studies. We refer the reader to \cite{WF} for more information on WF algorithm. WF is known to be optimal for such problems when Shannon capacity is used for the rate function. However, it is not optimal for the practical piecewise constant MCS rate function which we adopted.  We have done some simulations to verify if WF produces reasonable solutions. In the first set of simulations, we have solved the OPD problem with ZF-DBF by solving $\mathbf{\Pi}^\text{P1}_\text{OPD}$ followed by solving $\mathbf{\Pi}^\text{P2}_\text{OPD}$ using our approach. In the second set of simulations, we solve the OPD problem with ZF-DBF as follows: we use WF instead of solving $\mathbf{\Pi}^\text{P1}_\text{OPD}$ and then use the results to solve $\mathbf{\Pi}^\text{P2}_\text{OPD}$ using our approach. Fig.~5b shows the results for the case with $U=10$ UEs. We observe that while WF is sub-optimal, the gap between WF performance and our OPD approach is not large (around $10\%$ performance loss when $K=10$). This reveals that a combination of WF and $\mathbf{\Pi}^\text{P2}_\text{OPD}$ is sufficiently good in the case of ZF-DBF. We have conducted similar simulations for the case of N-DBF and had the same observation. These observations justify the use of WF in our online heuristic algorithms in Section \ref{sec:online}.

In summary, the offline study provides several important insights that help us design online heuristics for operation:
\begin{itemize}[leftmargin=*]
    \item Ignoring constraint ConB can yield significant performance over-estimation in practice.
    \item Good performance can be achieved even when $K$ is not chosen to be close to $U$. For example, with $U=10$ UEs, $K=6$ yields 92\% of the best performance.
    \item Both OPD and ZF-DBF can considerably improve the system performance, particularly when the BS can serve a large number of UEs within a PRB.
    \item Having more beams in the ABF codebooks can significantly boost the system performance, and it is more helpful to enlarge the codebook on the BS side than the UE side.
    \item EPD provides worse performance than OPD and less flexibility since the only ``knobs" are beam selection and user selection which makes fairness enforcement in an online heuristic difficult. 
\end{itemize}

\vspace{-10pt}
\section{Operation (Online RRM Study)} \label{sec:online}
This section is dedicated to the design of online RRM schemes, which aim to achieve high performance with low complexity. We only consider the OPD case since it performs better than the EPD cases and provides more flexibility. Unlike offline RRM studies that focus solely on improving system performance, our objective is to develop practical schemes that are easy to implement. We begin by introducing a low-complexity RRM benchmark scheme that can be used with ZF-DBF or without it (i.e., N-DBF). We then propose two novel and low-complexity heuristic RRM methods, one for ZF-DBF and one for N-DBF. In designing these methods, we leverage  Remark~\ref{remark} in Section~\ref{subsec:offline-sym}, which suggests selecting a single beam set for each MB and a single UE set for each RBL within an MB. Finally, we compare the performance and runtime of our heuristics and the low-complexity benchmarks. We also show the performance of the offline schemes  (Section~\ref{sec:offline}) to validate that our proposed heuristics strike an acceptable balance between performance and complexity.



In the following, we consider a particular BA period made of $N_o=100$ MBs (roughly half second) and describe the heuristics for an MB within this period. BA yields the preferred BS beam set $\mathcal{B}_p$. At the beginning of each MB, the inputs of the online schemes are $\mathcal{B}_p$, the $R_u$'s (the weights of each user) and the ECSI per RBL per user. Beam selection is about selecting a subset of $\mathcal{B}_p$ of size $L$ for each MB (recall that because of ConB, the beam set cannot change within an MB). Then, for each RBL, given the beam set, we perform \emph{in parallel}, user selection, DBF optionally, and PD.  At the end, we compute the rate seen by each user in the MB and update their weights accordingly.

\subsection{Low-complexity Benchmark} \label{subsec:online-bench}
In this section, we propose a low-complexity benchmark to perform beam and user selection based on a round-robin (RR) approach. Specifically, we list all the beams in 
 $\mathcal{B}_p$ and take $L$ beams in a RR fashion as a beam set for each of the $N_o$ MBs. Subsequently, given an MB and its beam set, the benchmark generates all possible UE sets of size $L$ (one UE per beam) to take into account the fact that multiple UEs might have the same preferred BS beam. We then  select one of these UE sets per RBL in a RR fashion inside the MB. In the case of the benchmark, the  beam and user selections only require the preferred BS beam of each UE so it can be done for all the $N_o$ MBs just after BA.

Next,  DBF (if available) and PD are performed for each RBL of an MB, given the selected UEs. Depending on the availability of DBF, there are two possible benchmarks: (i) $\mathscr{B}$(N-DBF), (ii) $\mathscr{B}$(ZF-DBF). To optimize PD in both (i) and (ii), similar to \cite{ECSIXiaomeng}, we adopt the water-filling (WF) algorithm discussed above.  Option (i) has a very low complexity since only limited computations are required per RBL for the WF algorithm and for computing the SINRs from the effective channels to select proper MCSs. Option (ii), on the other hand, requires additional computations per RBL to compute the ZF-DBF precoding.

\subsection{Proposed Heuristic Solutions}

We propose  RRM heuristics  for both ZF-DBF and N-DBF cases (called $\mathscr{H}$(N-DBF) and $\mathscr{H}$(ZF-DBF) resp.). Our proposed heuristics employ smarter beam and user selections than the ones used in the benchmarks. 
The building blocks of our  heuristics are illustrated in Fig.~\ref{Heuristic}. The beam selection (Algorithm~\ref{BeamS}) is the same for both heuristics and is called once per MB because of conB.   The user selection is called every RBL and is the same for both heuristics. PD differs for the two heuristics.   Specifically, given the beam set  $\ell_H$ for an MB, the following steps are performed in parallel for each RBL in an MB. First, we select UEs by Algorithm~\ref{US} for each RBL $q$. Given the UE set $z_H^q$ for each RBL, we perform  ZF-DBF per RBL for $\mathscr{H}$(ZF-DBF) (for $\mathscr{H}$(N-DBF) this step is skipped). Then, we perform PD for each RBL.  For $\mathscr{H}$(ZF-DBF), we found that performing PD with WF works well but this is not the case for $\mathscr{H}$(N-DBF). Hence, we  propose a new PD scheme that we call interference-aware WF (IAWF) (please see Algorithm~\ref{IAWF}) for that case. Note that there is no point applying IAWF to $\mathscr{H}$(ZF-DBF) since there is no interference after ZF precoding. We explain Algorithms~\ref{BeamS},~\ref{US} and~\ref{IAWF} next.

Our beam set selection and user selection are inspired by  the heuristic proposed for the case with single frequency subchannel in \cite{ECSIXiaomeng}. In that case, there is no need to do beam selection and the whole focus is on user selection. The heuristic in \cite{ECSIXiaomeng} computes a coefficient for each UE  and chooses the $L$ UEs (recall that $L=\min(K,|\mathcal{B}_p|)$) having the largest coefficients while ensuring that the preferred beams of the selected UEs are distinct.  We consider the general case where there are multiple frequency subchannels and hence, the need for beam selection and the enforcement of constraint ConB. Therefore, in our heuristics, in addition to defining UE coefficients, we propose the notion of beam coefficients. Defining both beam and UE coefficients enables us to perform beam and user selections in the multi-channel scenario.

\begin{figure}
    \centering
    \includegraphics[width=85mm]{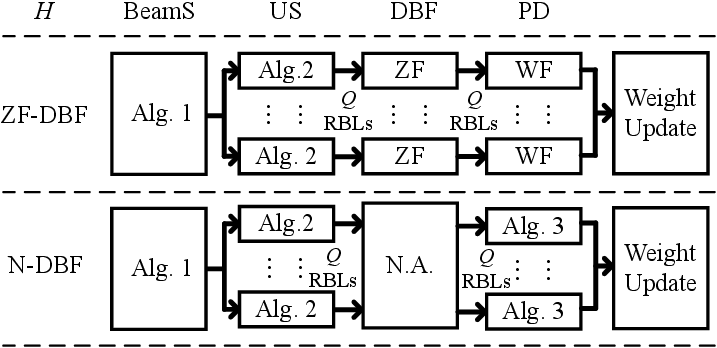}
    \caption{Building blocks of our online heuristics  for an MB (BeamS: beam selection, US: user selection, Alg.: Algorithm, N.A.: not available)}
    
    \label{Heuristic}
    \vspace{-10pt}
\end{figure}



To perform beam selection in an MB, we will first compute beam coefficients and then select the $L$ beams with the highest coefficients. Specifically, we first compute a coefficient per UE per RBL. To do so for RBL $q$, we assume that all UEs are selected in a PRB and distribute the power among UEs using WF with $|g_{q,u,u}^\text{eff}|^2$ as the gain and $1/R_u$ as the weight of UE $u$ for all $u \in \mathcal{U}$. Note that $g_{q,u,u}^\text{eff}$ is the effective signal channel gain of UE $u$ in RBL $q$ defined in Section~\ref{subsec:ECSI}. Then, for each UE, we compute its SNR (i.e., we neglect interference) and map it to a data rate based on the MCSs. The obtained data rate is used as the coefficient for the UE in RBL $q$. The UE coefficients in \cite{ECSIXiaomeng} are computed assuming that the full power is allocated to a user and Shannon capacity formula. The way we compute the coefficients creates a competition in the power domain among the UEs. Also, since we employ practical MCSs instead of the Shannon capacity formula, a UE coefficient will be zero if the SNR of the UE falls below the minimum MCS threshold. 

Next, we use the UE coefficients to compute the beam coefficient $c_b$ for each beam $b\in \mathcal{B}_p$ as $c_b=\sum_{q \in \mathcal{Q}}  c^q_{u^*_{b,q}}$, where $u^*_{b,q} = \arg\max_{\{u \text{ s.t. } b(u)=b\}} c^q_u$. In words, to compute beam coefficient $c_b$,  we first find, for each RBL $q$, the candidate UE $u^*_{b,q}$, that has the largest $c^q_u$ among all the UEs whose preferred beam is $b$. Then, we sum up such candidate UE coefficients over $q\in \mathcal{Q}$ to obtain $c_b$.
We then select the $L$ beams with the highest values of beam coefficients for that MB. Let $\ell_H$ denote this beam set. 

\begin{algorithm}
\small
\caption{Beam Selection at the beginning of an MB}
\label{BeamS}
\begin{algorithmic}[1]
    \For{$q=1:Q$} \algorithmiccomment{Compute UE coefficients}
        \State Assume all the UEs are selected in the $q$-th RBL. Distribute power to UEs using WF with $|g_{q,u,u}^\text{eff}|^2$ as the gain and $1/R_u$ as the weight of UE $u$ for all $u \in \mathcal{U}$. For all $u \in \mathcal{U}$, calculate the SNR of UE $u$ and map it to rate $r^q_u$ using the piece-wise constant rate function. 
        \State For all $u \in \mathcal{U}$, set $c^q_u$ to $r^q_u/R_u$. 
    \EndFor
    \For{$b \in \mathcal{B}_p$} \algorithmiccomment{Compute beam coefficients}
    \State Compute $c_b=\sum_{q \in \mathcal{Q}}  c^q_{u^*_{b,q}}$, where $u^*_{b,q} = \arg\max_{\{u \text{ s.t. } b(u)=b\}} c^q_u$.
    \EndFor
    \State Rank beams in $\mathcal{B}_p$ in a descending order of $c_b$. Select the first $L$ beams as $\ell_H$, the beam set for this MB. 
\end{algorithmic}
\end{algorithm}

User selection is performed by Algorithm~\ref{US} as follows in each RBL $q$ of an MB given $\ell_H$: for each selected beam $b^*$ in $\ell_H$, we select UE $u^*_{b^*,q}$ for RBL $q$ where $u^*_{b,q}$ was defined when defining beam coefficients above. Note that we do not select UEs with a coefficient of zero. 

\begin{algorithm}
\small
\caption{User Selection in RBL $q$ given $\ell_H$}
\label{US}
\begin{algorithmic}[1]
    \For{$b \in \ell_H$}
        \State Select UE $u$ with the highest $c^q_u$ (calculated in Algorithm~\ref{BeamS}) in $\{u:b(u)=b\}$ as a UE in the UE set $z_H^q$ for this RBL. UE $u$ \underline{will not be} selected if $c^q_u=0$. 
    \EndFor
\end{algorithmic}
\end{algorithm}

Finally, we present in Algorithm~\ref{IAWF}, the PD solution for  RBL $q$ for $\mathscr{H}$(N-DBF), which we call Interference-Aware WF (IAWF). We first estimate the interference of each UE in $z_H^q$ under the assumption in Algorithm~\ref{BeamS} line 2. We assume all UEs in $\mathcal{U}$ are selected and allocated power by the WF algorithm. Note that this part has been computed by Algorithm~\ref{BeamS} so we can use the calculated power directly. Then, we calculate the interference of UE $u$ in RBL $q$ as $I_u^q$ with the calculated power. We will ignore the interference if it is from a UE preferring the same BS beam as UE $u$. With $I_u^q$  $\forall u \in z_H^q$, we weigh their effective channel gain as $\frac{\sigma_\text{PRB}^2}{I^q_u+\sigma_\text{PRB}^2} |h_{q,u,u}(z_H^q)|^2$ and do WF algorithm based on the weighted effective channel gains. This will avoid distributing too much power to highly interfered UEs, which improves the overall rate performance.

\begin{algorithm}
\small
\caption{IAWF in RBL $q$ given $z_H^q$ for $\mathscr{H}$(N-DBF)}
\label{IAWF}
\begin{algorithmic}[1]
        \State Based on the assumption in Algorithm \ref{BeamS} line 2, calculate the total interference of UE $u$ from all other UEs except the ones sharing the same preferred BS beam as $I^q_u$ for all $u \in \mathcal{U}$.
        \State Perform PD for UEs in $z_H^q$ using WF with $\frac{\sigma_\text{PRB}^2}{I^q_u+\sigma_\text{PRB}^2} |h_{q,u,u}(z_H^q)|^2$ as the gain and $1/R_u$ as the weight of UE $u$. 
\end{algorithmic}
\end{algorithm}

\vspace{-5pt}

\subsection{Numerical Evaluations for Online RRM Study} \label{subsec:online-sym}

This section presents the simulation results for our heuristics and bechmarks using the same channel model, ABF codebooks, and BA approach as in our offline study. Since our  online algorithms are much faster than our offline solutions,  we increase the number of realizations to $Z=200$. All our results are for $B_b=32$, $B_u=8$, and $K=8$ in this section (all other parameters consistent with Table~\ref{Param}).

We compare the performance  of the heuristics and benchmarks in Fig.~\ref{r7} where we also show the results for the offline cases.
We compare the runtime  of the heuristics and benchmarks in Fig.~\ref{r8} for one MB. We would like to highlight that the  runtime of the offline solutions is in the order of multiple days which is much larger than the ones for the online algorithms.

Figs.~\ref{r7} and \ref{r8} show that $\mathscr{H}(\text{ZF-DBF})$ can achieve at least 92.3\% of the corresponding offline target performance and its performance is up to 338.6\% of that of $\mathscr{B}(\text{ZF-DBF})$ at the cost of 53.9\% higher maximum runtime. Meanwhile, $\mathscr{H}(\text{N-DBF})$ can achieve 90.2\% of its target performance, and its performance is 193.4\% of that of $\mathscr{B}(\text{N-DBF})$ with 124.9\% higher maximum runtime (recall that $\mathscr{B}(\text{N-DBF})$ has a very low complexity).  Fig.~\ref{r7} also shows the impact of DBF. We observe that ZF-DBF significantly improves the performance compared to N-DBF in our heuristics (by 47.3\%). The results confirm that our heuristics can strike a good balance between performance and runtime in both ZF-DBF and N-DBF scenarios.

\begin{figure}
    \centering
    \begin{subfigure}[b]{0.45\textwidth}
         \centering
         \includegraphics[width=70mm]{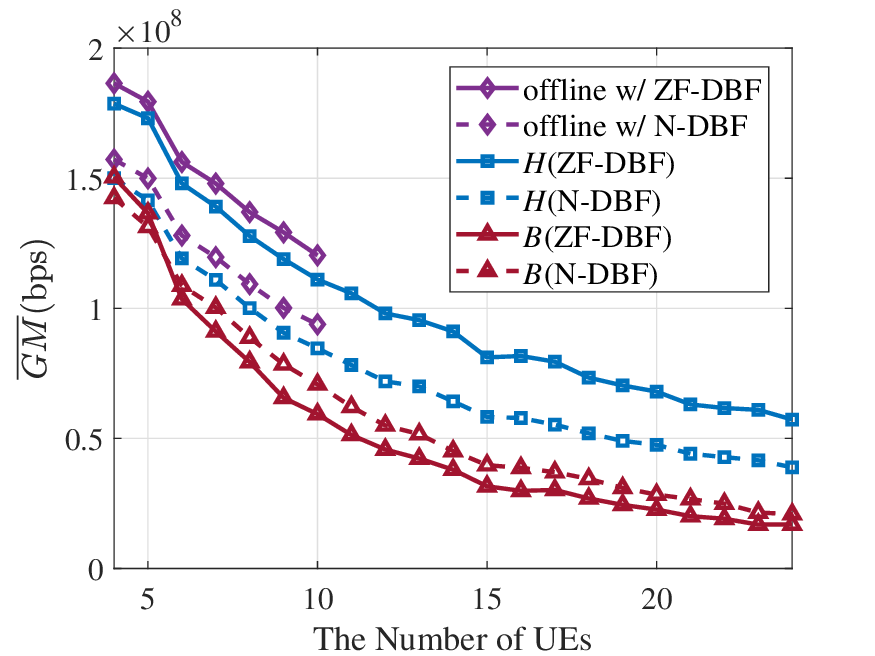}
         \caption{}
         \label{r7}
    \end{subfigure}
    \hfill
    \begin{subfigure}[b]{0.45\textwidth}
         \centering
         \includegraphics[width=70mm]{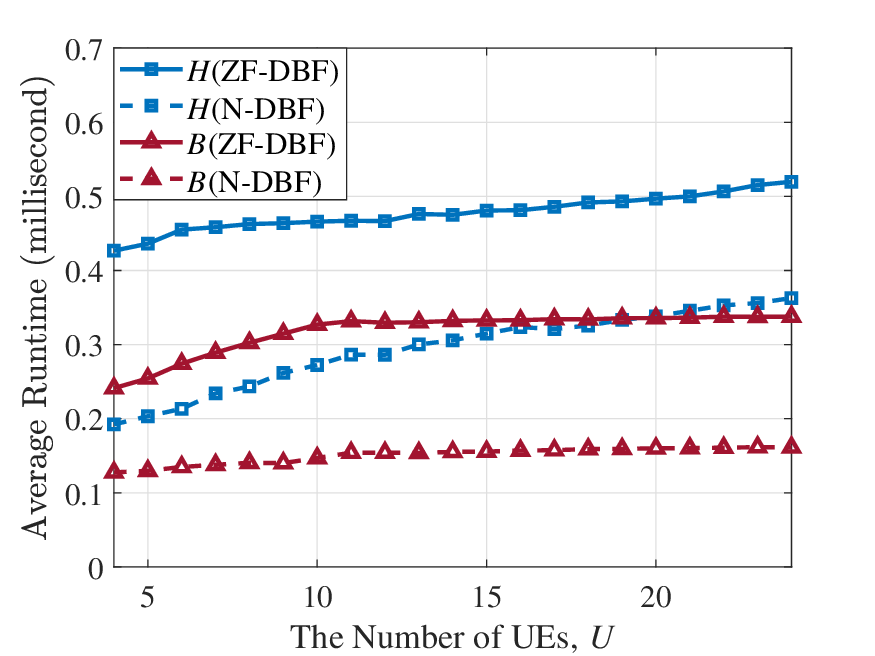}
         \caption{}
    \label{r8}
    \end{subfigure}
    \vspace{-5pt}
    \caption{Performance of the proposed heuristics and benchmarks as a function of $U$ with $K=8$, $N_b=128$, $N_u=16$, $B_b=32$, $B_u=4$.  (a) $\overline{\text{GM}}$, (b) average runtime per MB.}
    \vspace{-10pt}
\end{figure}




\section{Conclusion} \label{sec:conclusion}
In this paper, we have investigated downlink RRM in mmWave systems with codebook-based hybrid beamforming considering a practical multiple-channel scenario with a limited number of RF chains ($K<U$). This scenario  which has been mostly overlooked in the literature, brings several new RRM challenges due to the constraint on beam selection, that require new formulations and results. By formulating and solving an offline joint RRM optimization problem for different cases, we found that ignoring the beam selection constraint overestimates the performance by 20\%. We also show that  good performance can be achieved even when $K$ is smaller  than $U$. We also found that non-equal power distribution and ZF-DBF can improve the system performance by 22\% and 32\% compared to EPD and N-DBF, respectively. Another finding is that having one more pair of preferred beams can improve the performance by 22\% at the cost of CE overheads. Based on the insights gained from the offline study, we have proposed heuristic online algorithms that offer acceptable computational complexities and can reach 92\% of the performance targets obtained from the offline study. 

We have also investigated the use of Shannon capacity formula in our problems.  Using it in our offline optimization problem yields reasonable solutions because we have many degrees of freedom (i.e., variables) to play with. On the other hand, if we select beam and user sets in a non-careful way, online, waterfilling (WF) performs poorly. If we use WF on carefully chosen beam and user sets (i.e. so that the power per PRB is not spread too thin), then we can obtain good solutions.

\ifCLASSOPTIONcaptionsoff
  \newpage
\fi



%
\bibliographystyle{IEEEtran}
\bibliography{References.bib}
%








\end{document}